\documentclass[11pt]{article}
\pdfoutput=1
\usepackage{amsmath,amssymb,amsfonts,dsfont}
\usepackage{graphicx}
\usepackage{color}
\usepackage[citebordercolor={.8 .8 1},urlbordercolor ={.8 .8 1},pdfstartview=FitV]{hyperref}
\usepackage{latexsym}
\usepackage{verbatim}
\newcommand{\bea}{\begin{eqnarray}}
\newcommand{\eea}{\end{eqnarray}}
\newcommand{\be}{\begin{equation}}
\newcommand{\ee}{\end{equation}}
\newcommand{\ba}{\begin{array}}
\newcommand{\ea}{\end{array}}

\def\nn{\nonumber}

\def\p{\partial}
\def\eps{\epsilon}

\numberwithin{equation}{section}

\def\cQ{\mathcal{Q}}
\def\cL{\mathcal{L}}
\def\cW{\mathcal{W}}

\def\sdelta{\slash\hspace{-6pt}\delta}

\definecolor{Wei}{rgb}{0.65,0.0,0}
\definecolor{Geo}{rgb}{0.1,0,0.75}

\newcommand{\stress}{\mathcal{L}}
\newcommand{\stressbar}{\bar{\mathcal{L}}}
\newcommand{\tstress}{\tilde{\mathcal{L}}}
\newcommand{\tstressbar}{\bar{\tilde{\mathcal{L}}}}
\newcommand{\wcharge}{\mathcal{W}}
\newcommand{\twcharge}{\tilde{\mathcal{W}}}
\newcommand{\wchargebar}{\bar{\mathcal{W}}}

\newcommand{\fudge}{\gamma}

\numberwithin{equation}{section}

\begin{document}
\begin{center}

\vspace{1cm}

 { \Large {\bf Observables and Microscopic Entropy\\ of Higher Spin Black Holes}}

\vspace{1cm}

Geoffrey Comp\`ere$^{\clubsuit\Diamond}$, Juan I. Jottar$^\spadesuit$ and Wei Song$^{\clubsuit\heartsuit}$

\vspace{0.8cm}

{\it  $\clubsuit$ Center for the Fundamental Laws of Nature, Harvard University,\\
Cambridge, MA 02138, USA}\\\vspace{0.2cm}

{\it  $\Diamond$ Physique Th\'eorique et Math\'ematique, Universit\'e Libre de Bruxelles,\\
 Bruxelles, Belgium}\\\vspace{0.2cm}

{\it  $^\spadesuit$ Institute for Theoretical Physics, University of Amsterdam,\\
1090 GL Amsterdam, The Netherlands}\\\vspace{0.2cm}

{\it  $^\heartsuit$ Department of Physics, Princeton University,
Princeton, NJ 08544, USA}\\\vspace{0.2cm}

\vspace{1.5cm}

\end{center}

\begin{abstract}
In the context of recently proposed holographic dualities between higher spin theories in AdS$_{3}$ and $(1+1)$-dimensional CFTs with $\mathcal{W}$ symmetry algebras, we revisit the definition of higher spin black hole thermodynamics and the dictionary between bulk fields and dual CFT operators. We build a canonical formalism based on three ingredients: a gauge-invariant definition of conserved charges and chemical potentials in the presence of higher spin black holes, a canonical definition of entropy in the bulk, and a bulk-to-boundary dictionary aligned with the asymptotic symmetry algebra. We show that our canonical formalism shares the same formal structure as the so-called holomorphic formalism, but differs in the definition of charges and chemical potentials and in the bulk-to-boundary dictionary. Most importantly, we show that it admits a consistent CFT interpretation. We discuss the spin-2 and spin-3 cases in detail and generalize our construction to theories based on the $\text{hs}[\lambda]$ algebra, and on the $sl(N,\mathds{R})$ algebra for any choice of $sl(2,\mathds{R})$ embedding.

\end{abstract}
\thispagestyle{empty}

\pagebreak
\setcounter{tocdepth}{3}

\tableofcontents

\thispagestyle{empty}

\section{Introduction and summary of results}

The recent surge in the study of holographic dualities involving higher spin theories has allowed us to better understand the relationship between gravity and quantum field theories in different regimes of parameters. Starting with the proposal of Klebanov and Polyakov \cite{Klebanov:2002ja} linking the Fradkin-Vasiliev theories in AdS$_{4}$ \cite{Fradkin:1987ks,Fradkin:1986qy} and $O(N)$ vector models in three dimensions, it has become increasingly clear that higher spin theories in AdS are dual to ``simple" CFTs, in the sense that the structure of correlators is strongly constrained by the symmetries along the lines discussed e.g. in \cite{Maldacena:2012sf,Maldacena:2011jn}.  In fact, dualities involving higher spin fields are examples of weak-coupling/weak-coupling dualities, in contrast with the complicated strongly-coupled field theories which are usually studied in the context of AdS/CFT via standard classical gravity duals \cite{Maldacena:1997re,Witten:1998qj}. Ultimately, we expect the study of these complementary regimes of the correspondence to improve our understanding of quantum gravity and gauge theories alike.

It is in general difficult to define physical observables in a higher spin gauge theory that contains a gravitational sector. One obstacle stems from the fact that additional higher spin gauge transformations blur the notion of geometry: curvature singularities and geodesics are not gauge-invariant quantities, for example. Furthermore, Vasiliev's theory \cite{Vasiliev:1990en,Vasiliev:1992av} does not have a known action formulation, which complicates the definition of an ADM-type energy (see \cite{Boulanger:2011dd,Sezgin:2011hq} for recent partial proposals, however). The latter difficulty can nonetheless be circumvented in the pure higher spin theory in three dimensions \cite{Blencowe:1988gj,Bergshoeff:1989ns}, making the lower-dimensional setup a promising arena to explore this class of holographic dualities. Motivated by this fact, in the present paper we will propose gauge-invariant definitions of conserved charges in three-dimensional higher spin theories which agree with those obtained via canonical methods, and moreover show that these definitions lead to a natural CFT interpretation of the higher spin black hole thermodynamics.

It was conjectured \cite{Gaberdiel:2010pz} that a version of Vasiliev's higher spin theory on AdS$_3$ \cite{Vasiliev:1995dn,Vasiliev:1996hn} is holographically dual to certain coset minimal model CFTs in the large-$N$ limit. A consistent truncation is possible where the matter sector decouples, and the pure higher spin theory then reduces to a Chern-Simons theory based on the Lie algebra $\text{hs}[\lambda] \oplus \text{hs}[\lambda]\,$, or its further truncations to $sl(N,\mathds{R}) \oplus sl(N,\mathds{R})$ when $\lambda = \pm N$ (with $N$ an integer). The dual $(1+1)$-dimensional CFTs enjoy (two copies of) $\cW_\infty[\lambda]$ and $\cW_N$ symmetry, respectively. Notably, these Chern-Simons theories have been shown to admit black hole solutions carrying higher spin charges \cite{Gutperle:2011kf,Kraus:2011ds,Ammon:2012wc}. However, as we will now review, the current status of the higher spin black hole thermodynamics and microscopic entropy counting is in some ways still unsatisfactory.

From the bulk point of view, two formalisms have been proposed in order to define the higher spin thermodynamics, which however disagree on their definition of entropy as well as on the physical observables. The so-called ``holomorphic formalism" was originally developed in \cite{Gutperle:2011kf,Ammon:2011nk}. There, it was noted that higher spin black holes violate the Brown-Henneaux boundary conditions \cite{Brown:1986nw,Banados:1994tn,Coussaert:1995zp}. The offending terms are induced by two deformation parameters $\mu_j,\, \bar\mu_j$ for each spin-$j$ field, which are closely related to chemical potentials in the Euclidean formulation (which are the thermodynamic conjugate of the higher spin charges). The physical charges (or more generally the observables) in the holomorphic formalism are obtained from the gauge connections in the same way as in standard asymptotically AdS$_3$ solutions, namely, when $\mu_j=\bar\mu_j=0\,$. In Euclidean signature, requiring that the holonomy around the thermal circle is trivial gives the relation between the charges and their conjugate chemical potentials. The entropy can be then defined by integrating the first law of thermodynamics, and the partition function in the saddle point approximation can be obtained as a Legendre transform of the entropy. For additional developments, see \cite{Ammon:2011ua,Kraus:2012uf,David:2012iu,Kraus:2013esi,Gaberdiel:2013jca}.

An alternative formalism was developed using canonical methods along the lines of \cite{Regge:1974zd,Abbott:1981ff,Brown:1986ed,Lee:1990nz,Barnich:2001jy}. The canonical entropy, which turns out to disagree with the proposed holomorphic definition, can be obtained in various ways. First, it can be derived using Wald's formula \cite{Wald:1993nt} after formulating Chern-Simons higher spin theory in the metric formalism \cite{Campoleoni:2012hp,Banados:2012ue,Perez:2013xi}, at least perturbatively in the higher spin sources. In \cite{Banados:2012ue,deBoer:2013gz}, boundary terms that make the variational principle well-defined for fixed $\mu_j,\bar\mu_j$ were constructed and the canonical entropy was defined as the Legendre transform of the resulting free energy. Recently, the canonical entropy was also found as the thermal limit of entanglement entropy  \cite{Ammon:2013hba,deBoer:2013vca,Caputa:2013eka} and in particular obtained from a generalization of the conical deficit method \cite{Ammon:2013hba}. The discrepancies in the notion of entropy can be traced back to different definitions of conserved charges and chemical potentials, an issue that has remained somewhat controversial. In \cite{Perez:2012cf,Perez:2013xi} the energy was defined as the conserved charge associated with the timelike Killing vector of the asymptotically undeformed AdS$_3$ metric. Another definition based on the free energy and the Euclidean time periodicity was proposed in \cite{Banados:2012ue}.  A boundary stress-tensor was defined in \cite{deBoer:2013gz} which led to an energy agreeing with \cite{Perez:2012cf,Perez:2013xi} after additional considerations on the variational principle. A different proposal \cite{Compere:2013aa}, that we use as a guiding line here, is to associate the energy with the zero mode generators of the asymptotic symmetry algebra in the presence of the deformation. While the definition of entropy is natural from the canonical perspective, it is clear that a universal definition of physical observables that is well motivated from both the bulk and CFT perspectives is lacking.

Before stating our results, let us comment on the interpretation of higher spin black holes in terms of the conjectured dual CFT. On the one hand, in the holomorphic formalism, higher spin black holes are interpreted as states in a CFT deformed by an irrelevant operator. The states are then proposed to be counted using conformal perturbation theory around the undeformed CFT \cite{Gutperle:2011kf}. Since the asymptotic symmetry algebra at $\mu_j=\bar\mu_j=0$ consists of the direct sum of holomorphic and anti-holomorphic $\cW$ algebras \cite{Henneaux:2010xg,Campoleoni:2010zq,Gaberdiel:2011wb,Campoleoni:2011hg}, the partition function in the saddle point approximation is holomorphically factorized. Using as  bulk-to-boundary dictionary the undeformed AdS/CFT dictionary (see \cite{Kraus:2006wn} for a review), the CFT calculation reproduces the macroscopic calculation in the holomorphic formalism. It is however striking that the resulting entropy disagrees with the canonical definitions. We will argue that this mismatch points to the need for a refinement of the bulk-to-boundary dictionary in the holomorphic formalism. On the other hand, the connection between the canonical formalism and the dual CFT interpretation is so far unclear. To sum it up loosely, it appears that the holomorphic formalism is more friendly with a CFT interpretation while the canonical formalism is more natural from the gravity (bulk) perspective. Since these two formalisms are associated with different partition functions, it is crucial to understand which one is the preferred formulation from the gravitational perspective, while achieving a clear understanding of the same picture from the dual CFT perspective.

Our main objective in this paper is to propose a unified formalism to define the thermodynamics of higher spin black holes, consistent both with canonical methods and with a microscopic CFT interpretation. The first pillar of our construction is the definition of conserved charges $\tilde Q_j$ and their dual sources $\tilde \alpha_j$, for each spin-$j$ field (the definitions are similar in the barred sector and will be omitted here). To distinguish these from quantities in the undeformed theory, we put a tilde on these new observables. We will sometimes refer to these variables as ``the tilded variables''. We will denominate $\mu_j$ solely as deformation parameters, while keeping the terminology of chemical potentials for $\tilde \mu_j$ and boundary sources for $\tilde \alpha_j\,$. We define the conserved charges from the holonomies of the reduced gauge connection\footnote{By reduced gauge connection we mean one where the gauge freedom has been employed to gauge away the dependence on the bulk radial coordinate, leaving a connection that depends on the boundary coordinates $z,\bar z$ only.} $a=a_z dz+a_{\bar z}d{\bar z}$ around the spatial boundary circle parameterized by $\varphi = (z+\bar z)/2 \sim \varphi +2\pi\,$. For constant $a$ the canonical charges are defined as
\bea
\tilde Q_j=k_{cs}\text{Tr}\bigl[V^j_{j-1}a_\varphi\bigr] = k_{cs} N_j \text{Tr}\left[a_\varphi b_{j-1}(a_\varphi)\right],
\eea
where $k_{cs}$ is the level of the Chern-Simons theory,  $V^j_{j-1}$ is the highest-weight generator for the $sl(2,\mathds{R})$ spin-$(j-1)$ multiplet, $b_{j-1}=a_\varphi^{j-1}+\cdots $ is a  polynomial in $a_\varphi$ and $N_j$ is a normalization factor; both $b_{j-1}$ and $N_j$ are uniquely fixed by matching the second with the first definition. The first definition is only valid in the highest-weight gauge for $a_\varphi\,$, while the second definition is manifestly invariant under any regular large gauge transformation, since the latter expression only depends on holonomies around the boundary $\varphi$ circle. For the spin-3 case, it was found in \cite{Compere:2013aa} using perturbative canonical and integrability methods that the asymptotic symmetry algebra that describes Dirichlet boundary conditions in the principal embedding consists of two copies of the $\cW_3$ algebra, even in the presence of $\mu_3,\bar\mu_3$ deformations. It turns out that the gauge-invariant definition exactly matches with the zero modes of the $\cW_3 \oplus \cW_3$ algebra obtained in perturbation theory in \cite{Compere:2013aa}. (Note that the non-zero modes cannot be written as holonomies since they correspond to local boundary excitations).

The second guiding principle behind our formalism is the canonical definition of entropy. We propose to define the entropy as the canonical charge associated with, as gauge parameter, the component of the connection along the thermal Euclidean circle. Much like Wald's entropy formula \cite{Wald:1993nt}, this definition is independent of boundary terms in the action and relies on fundamental properties of the black hole only. Furthermore, we show that this definition is consistent with the first law of thermodynamics and recover the general formula
\bea
S = - 2\pi i k_{cs} \text{Tr}\Bigl[\left(a_{z}  +a_{\bar{z}}\right) \left(\tau a_z + \bar \tau a_{\bar z}\right) - \left(\bar{a}_{z} + \bar{a}_{\bar{z}}\right) \left(\tau \bar{a}_z + \bar{\tau} \bar{a}_{\bar z}\right)\Bigr],
\eea

\noindent  first derived in \cite{deBoer:2013gz}, which gives the entropy of higher spin black holes in terms of the reduced connection. 

The third and last pillar in our construction is the bulk-to-boundary dictionary. We align the dictionary to the asymptotic symmetry algebra analysis, as usually done in holographic correspondences \cite{Strominger:1997eq}. In the spin-3 case, the asymptotic symmetry algebra was computed in \cite{Compere:2013aa} and led to identify the tilded variables as the dual CFT variables.  In this paper we provide further evidence for this proposal for the theory based on the general hs$[\lambda]$ algebra, and in particular the $sl(N,\mathds{R})$ theory with any choice of  $sl(2,\mathds{R})$ embedding. We will obtain that the partition function for higher spin black holes in the saddle point approximation can be written as\footnote{As written here, this formula is appropriate for the theory based on the hs$[\lambda]$ algebra or the $sl(N,\mathds{R})$ theory in the principal embedding (in which the sum terminates at $j=N$). Similar expressions apply in other embeddings.}
\be
\ln Z_{\text{bulk}}=-2\pi i\sum_{j } (j-1)\tilde{\alpha}_{j} \tilde{Q}_{j}
\ee
(with a similar contribution from the barred sector) while the CFT partition function computed in \cite{Kraus:2011ds,Gaberdiel:2012yb} is given by
\be
 \ln Z_{\text{CFT}}=\ln \text{Tr}_{\mathcal{H}}\left[ e^{2\pi i\sum_{j } \alpha_{j;\text{CFT}} \hat{Q}_{j;\text{CFT}}} \right]
 =-2\pi i\sum_{j } (j-1){\alpha}_{j;\text{CFT}} {Q}_{j;\text{CFT}}
 \ee
where $\text{Tr}_{\mathcal{H}}$ denotes a trace over the Hilbert space of the CFT, $\hat{Q}_{j;\text{CFT}}$ is the zero mode of the dimension-$j$ current, $\alpha_{j;\text{CFT}}$ its conjugate source, and $Q_{j;\text{CFT}}\,$ denotes the expectation value of the corresponding conserved charge in the thermodynamic limit. We will argue in favor of the following bulk-to-boundary dictionary:
\bea
\tilde{\alpha}\rightarrow \alpha_{j;\text{CFT}}\,,\qquad \qquad \tilde{Q}_j\rightarrow{Q}_{j;\text{CFT}}\,.
\eea
It is then obvious that a CFT calculation would reproduce the canonical formalism calculation. We will illustrate the exact correspondence between bulk and boundary partition functions in the example of the $\text{hs}[\lambda]$ black hole studied in \cite{Kraus:2011ds,Gaberdiel:2012yb}.

The main idea underlying our construction of observables is that the conformal $\cW$-symmetry structure prevails in the presence of non-trivial deformation parameters. To a certain extent, this structure can be made more explicit using field redefinitions and gauge transformations; while technically hard to achieve for higher spin gauge theories, this is straightforward in the spin-2 case. There, the Virasoro Ward identities appear as the equations of motion when one imposes Dirichlet boundary conditions with fixed $\mu_2,\bar \mu_2$ deformation parameters \cite{Banados:2004nr}. We show that a combined field redefinition and gauge transformation exists that maps the system to the usual Brown-Henneaux boundary conditions, where the standard thermodynamics and microscopic counting apply \cite{Brown:1986nw,Strominger:1997eq}. If instead one uses the holomorphic formalism to compute the BTZ black hole entropy, one finds a result which contradicts the standard bulk entropy given by the area of the horizon divided by $4G_{3}$ (namely the Bekenstein-Hawking entropy). This example illustrates the conflict between the holomorphic formalism in its current formulation and the canonical entropy, and the importance of tilded variables to uncover the conformal symmetry preserved after the deformation.\footnote{We thank P. Kraus for pointing out that in the spin-2 case the holomorphic entropy formula can be reconciled with the Bekenstein-Hawking entropy at the price of changing the holonomies of the BTZ solution. This suggests that there might be a map between the two formalisms that remains to be fully understood.}

The discussion of higher spin algebras requires more care. In the principal embedding, the addition of deformation parameters corresponds to the addition of irrelevant operators to the original dual CFT, as opposed to marginal ones as in the spin-2 case. Nevertheless, it was shown in  \cite{Compere:2013aa} that there is a $\cW_3$ structure in perturbation theory in the deformation parameters. A natural concern is whether or not the analysis of \cite{Compere:2013aa} has a finite radius of convergence. Here we will provide further evidence that it does, by deriving the vacua of the $sl(3,\mathds{R})$ theory in the principal embedding at finite values of the deformation parameters. It is natural to define vacua as solutions which have trivial holonomy around the boundary $\varphi$ circle, much like the global AdS$_{3}$ solution of the standard (spin-2) gravity theory. We will show that two classes of vacua exist for deformation parameters $\mu_3,\bar \mu_3$ below a critical value, and prove that they admit the same number of symmetries as the original AdS$_3$ vacuum.

Extrapolating to arbitrary gauge algebras, the final picture that emerges from our analysis is the following. Since conceptually nothing changes for higher spins, we expect that the tilded charges will be identified with the zero modes of the asymptotic symmetry generators for any gauge algebra and any embedding.  Given a $\mu_j, \bar \mu_j$ deformation of the original CFT, appropriate variables (the tilded variables) will exist such that the conformal $\cW$ symmetry is preserved. The entropy of thermal states could then be counted by relating the behavior of the partition function at large and small temperature (generalizing Cardy's analysis \cite{Cardy:1986ie} for higher spin algebras) in terms of these appropriate tilded variables. By construction, this computation is exactly the one performed in \cite{Gaberdiel:2012yb} and it has exactly the same form as the entropy obtained in the holomorphic formulation \cite{Gutperle:2011kf,Ammon:2011nk}, but written in tilded variables instead. Since the bulk-to-boundary dictionary is aligned with the asymptotic symmetry algebra, the CFT zero-modes are the conserved charges carried by the black hole, and the bulk entropy is reproduced by the CFT counting. We therefore reconcile the holomorphic and canonical formalisms by keeping the formal structure of the holomorphic theory, but changing the holographic bulk-boundary dictionary to tilded variables. As a non-trivial check of these claims, we will show explicitly for the spin-3 case how to map the holomorphic entropy of the generic rotating black hole of \cite{Gutperle:2011kf} to the canonical entropy of \cite{deBoer:2013gz} once the variables in the holomorphic formalism are replaced by tilded variables.

The layout of the paper is as follows. In section \ref{sec2} we consider the $sl(2,\mathds{R})\oplus sl(2,\mathds{R})$ Chern-Simons theory with modified Brown-Henneaux boundary conditions where the boundary metric is deformed. We compute the asymptotic symmetry algebra, derive the thermodynamics, and give a microscopic counting of the entropy. It provides a technically straightforward summary of our methodology, that we further develop for the spin-3 case (i.e. the $sl(3,\mathds{R})$ algebra in the principal embedding) in section \ref{sec:spin3}. There, we present our definition of conserved charges and chemical potentials, and prove that the canonical entropy takes the form of the holomorphic entropy in tilded variables. We also discuss the maximally-symmetric vacua at finite $\mu_3,\bar \mu_3$ deformations. We extend our discussion to the more general $sl(N,\mathds{R})$ and $\text{hs}[\lambda]$ Chern-Simons theories in section \ref{sec:generalization} and show explicitly the matching to CFT calculations done in \cite{Gaberdiel:2012yb}. Our conventions are summarized in the appendices. We keep our conventions for $\text{hs}[\lambda]$ parameterized in terms of two arbitrary c-numbers $(q,\fudge)$, to allow for easier comparison with other references in the literature.

\section{Warm-up: $sl(2,\mathds{R})$ with deformations}
\label{sec2}
Before studying the thermodynamics of higher spin black holes in the presence of deformations by higher spin currents, it is instructive to consider a closely related problem in the pure gravity theory, corresponding to $SL(2,\mathds{R})\times SL(2,\mathds{R})$ gauge group \cite{Achucarro:1987vz,Witten:1988hc}. We start with the Lorentzian theory, where the boundary is topologically a cylinder, and employ light-cone coordinates $x^{\pm} = t/\ell \pm\varphi\,$, where $\varphi$ is the angular variable with period $2\pi\,$. Our conventions for the $sl(2,\mathds{R})$ algebra can be found in appendix \ref{subsec:sl2 conventions}.

 As pointed out in \cite{Banados:1998gg}, under the standard Brown-Henneaux boundary conditions \cite{Brown:1986nw} the line element
\begin{equation}\label{AAdS3 metric}
ds^{2} = \ell^{2}\left[d\rho^{2} +\frac{\stress}{k}\,(dx^{+})^{2} + \frac{\stressbar}{k}\,(dx^{-})^2- \left(e^{2\rho} +\frac{\stress\stressbar}{k^{2}} e^{-2\rho}\right)dx^{+}dx^{-} \right]
\end{equation}
\noindent with $k = \ell/(4G_{3})\,$  represents the whole space of asymptotically AdS$_{3}$ 
solutions with a flat boundary metric at $\rho \to \infty$, and $\stress$ and $\stressbar$ are seen to correspond to the stress tensor in the dual CFT (see \cite{Kraus:2006wn} for a review of the AdS$_{3}$/CFT$_{2}$ correspondence). In this light, Einstein's equations are equivalent to the holomorphicity properties
\bea
\p_{-} \stress = 0\,,\qquad \p_{+} \stressbar = 0\,,
\eea

\noindent encoding the conservation of the stress tensor. In the Chern-Simons formulation of the theory, the metric \eqref{AAdS3 metric} corresponds to the gauge connections
\begin{align}\label{AAdS solution}
A={}&
 \left(e^{\rho}L_{1} -e^{-\rho}\frac{\stress}{k}L_{-1}\right)dx^+ + L_{0}\,d\rho \,,
 \\
 \bar{A} ={}&
  -\left(e^{\rho}L_{-1} - e^{-\rho}\frac{\stressbar}{k}L_{1}\right)dx^{-}-L_{0}\,d\rho\,.
\end{align}

Let us now modify the theory by adding spin-2 deformation parameters $\mu_2(x^\pm)$, $\bar\mu_2(x^\pm)$, i.e. we change the metric on which the field theory is defined as
\begin{equation}
ds^{2}_{(0)} =  -\ell^{2}\left(dx^{+} + \mu_{2} dx^{-}\right)\left(dx^{-} + \bar{\mu}_{2}dx^{+}\right),\label{bndmetric}
\end{equation}
where we restrict $-1 < \mu_{2} < 1$ and $-1 < \bar{\mu}_{2}<1$ in order to preserve the boundary light-cone structure. In the Chern-Simons formulation, this deformation amounts to turning on the $A_{-}$ and $\bar{A}_{+}$ components of the connections, which now read
 \begin{align}
A={}&
 \left(e^{\rho}L_{1} -e^{-\rho}\frac{\stress}{k}L_{-1}\right)dx^+ + L_{0}\,d\rho
 \nn \\
 &+\left(   e^{\rho}\mu_{2}\,L_{1} +e^{-\rho}\left( - \mu_{2}\frac{\stress}{k}+\frac{1}{2}\partial_{+}^2 \mu_{2}\right)\,L_{-1} - \partial_{+}\mu_{2}\, L_0 \right) dx^- ,\label{Amm} \\
 \bar{A} ={}&
  -\left(e^{\rho}L_{-1} - e^{-\rho}\frac{\stressbar}{k}L_{1}\right)dx^{-}-L_{0}\,d\rho \nn \\
  &+ \left(    -e^{\rho}\bar{\mu}_{2}\,L_{-1} +e^{-\rho}\left(\bar{\mu}_{2}\frac{\stressbar}{k}-\frac{1}{2}\partial_{-}^2 \bar{\mu}_{2} \right)\,L_{1} + \partial_{-}\bar{\mu}_{2}\, L_0\right)dx^+ .
\end{align}

\noindent The equations of motion (flatness of the gauge connections) further imply
\begin{align}
\partial_{-}\stress ={}& \mu_{2}\,\partial_{+}\stress + 2\stress \partial_{+}\mu_{2} -\frac{k}{2}\partial^{3}_{+}\mu_{2}\,,
\nn\\
\partial_{+}\stressbar ={}& \bar{\mu}_{2}\,\partial_{-}\stressbar + 2\stressbar \partial_{-}\bar{\mu}_{2} -\frac{k}{2}\partial^{3}_{-}\bar{\mu}_{2}\,,\label{VW}
\end{align}

\noindent which we recognize as the Virasoro Ward identities associated with a deformation $\int d^{2}x\big(\mu_{2}\stress$ $- \bar{\mu}_{2}\stressbar\big)$ of the boundary field theory action.

Several natural questions arise: defining boundary conditions with fixed deformation parameters, what are the conserved charges and the asymptotic symmetry algebra? Moreover, can one give a microscopic interpretation of the BTZ black hole entropy? In order to answer these questions, we propose to perform a field redefinition in order to map the system to the standard Brown-Henneaux form. We will first discuss explicitly the case of constant deformation parameters which is the main point of interest in the context of stationary higher spin black holes. We will then briefly discuss the case of arbitrary deformations.

First, we would like to undo the twisting of the light-cone generated by the sources $(\mu_{2},\bar{\mu}_{2})$, while preserving the identifications of $x^{\pm}$ under $\varphi \to \varphi + 2\pi\,$, so that the angular coordinate of the boundary cylinder remains canonically normalized. We are then led to consider the following change of coordinates:
\begin{equation}
\tilde{x}^{+} = \frac{x^{+}+\mu_{2}\,x^{-}}{1-\mu_{2}}\,,\qquad \tilde{x}^{-} =  \frac{x^{-}+\bar{\mu}_{2}\,x^{+}}{1-\bar{\mu}_{2}}\,,\qquad\tilde{\rho} =\rho + \frac{1}{2}\ln\Bigl[\left(1-\mu_{2}\right)\left(1-\bar{\mu}_{2}\right)\Bigr],
\end{equation}
\noindent followed by a gauge transformation
\begin{equation}
\tilde A = e^{-\Delta L_0} A e^{\Delta L_0}, \qquad \bar{\tilde A} = e^{-\Delta L_0} \bar{A} e^{\Delta L_0} \label{chgtv}
\qquad\text{with}\quad \Delta = \frac{1}{2} \ln\left( \frac{1-\bar \mu_2}{1-\mu_2}\right).
\end{equation}
\noindent Notice that this gauge transformation belongs to the diagonal subgroup of $SL(2,\mathds{R})\times SL(2,\mathds{R})$ and it is therefore a rotation of the local Lorentz frame (in particular, it does not change the line element). The gauge connections are now
\begin{align}\label{tilde AAdS solution}
\tilde A={}&
 \left(e^{\tilde{\rho}}L_{1} -e^{-\tilde{\rho}}\frac{\tstress}{k}L_{-1}\right)d\tilde{x}^{+} + L_{0}\,d\tilde{\rho} \,,
 \\
 \bar{\tilde A} ={}&
  -\left(e^{\tilde{\rho}}L_{-1} - e^{-\tilde{\rho}}\frac{\tstressbar}{k}L_{1}\right)d\tilde{x}^{-}-L_{0}\,d\tilde{\rho}\,,
\end{align}
\noindent and the associated metric reads
\begin{equation}\label{AAdS3 metrictilded}
ds^{2} = \ell^{2}\left[d\tilde{\rho}^{2} +\frac{\tstress}{k}\,(d\tilde{x}^{+})^{2} + \frac{\tstressbar}{k}\,(d\tilde{x}^{-})^2- \left(e^{2\tilde{\rho}} +\frac{\tstress\tstressbar}{k^{2}} e^{-2\tilde{\rho}}\right)d\tilde{x}^{+}d\tilde{x}^{-} \right],
\end{equation}

\noindent where we defined the tilded quantities
\begin{equation}
\tstress = \left(1-\mu_{2}\right)^2\stress \,,\qquad \tstressbar = \left(1-\bar{\mu}_{2}\right)^2\stressbar \,.
\end{equation}
\noindent Therefore, we find that the metric (as well as the gauge connections) goes back to its original form, but written in terms of $\tstress$, $\tstressbar$ and in terms of tilded coordinates. One can then repeat the Brown-Henneaux analysis and obtain that the asymptotic symmetry algebra consists of two copies of the Virasoro algebra with central charge $c =6k$ as usual. Constant deformations do not therefore modify the asymptotic symmetry algebra nor break any symmetries. They merely transform the stress-tensor and boundary coordinates in a non-trivial way.

Notice that the zero modes of the Virasoro algebra can be defined from the $A_{\varphi}$, $\bar{A}_{\varphi}$ components of the connection as
\begin{equation}\label{new stress}
\tilde{\stress} = \frac{k}{2}\text{Tr}\bigl[\tilde A_{\tilde{\varphi}}^2\bigr]= \frac{k}{2}\text{Tr}\bigl[A_{\varphi}^2\bigr]\,,\qquad \bar{\tilde{\stress}} =  \frac{k}{2}\text{Tr}\big[\bar{\tilde A}_{\tilde{\varphi}}^2\bigr]= \frac{k}{2}\text{Tr}\bigl[\bar{A}_{\varphi}^2\bigr],
\end{equation}
where $\tilde A,\bar{\tilde A}$ can be found in \eqref{tilde AAdS solution}, while $A,\bar A$ are given in \eqref{Amm}. We see that this definition of the zero modes is in fact invariant under the coordinate transformation and gauge transformation we performed, and more generally under any gauge transformation that preserves the periodicity of the boundary spatial circle.

Among the metrics of the form \eqref{AAdS3 metrictilded}, a solution of particular interest is the BTZ black hole \cite{Banados:1992wn} with mass $M$ and angular momentum $J\,$, obtained for constant $\tilde{\stress}$, $\bar{\tilde{\stress}}$,  given as
\begin{equation}
\begin{aligned}\label{BTZ L and Lbar}
\tilde{\stress}_{\text{BTZ}} &= \frac{1}{2}\left(M\ell  - J\right) ,
\qquad
\bar{\tilde{\stress}}_{\text{BTZ}} = \frac{1}{2}\left(M\ell  + J\right) .
\end{aligned}
\end{equation}
\noindent In order to discuss its thermodynamics, let us now pass to Euclidean signature by a Wick rotation $x^{+} \to z$ and $x^{-} \to -\bar{z}\,$, with the identifications
\begin{equation}\label{z periodicities}
z \simeq z+ 2\pi \simeq z +2\pi \tau\,,
\end{equation}
where $\tau$ is the modular parameter of the torus on which the undeformed dual CFT is defined.

\noindent As usual, we can use the gauge freedom to isolate the radial dependence by defining the reduced connection $a(z,\bar{z})$ (and similarly for $\bar{a}$) as
\be\label{radial gauge def}
A=b^{-1} a b+b^{-1}db\,,\qquad \text{with}\quad b=e^{ \rho L_0}\,.
\ee
From the equations of motion and the $sl(2,\mathds{R})$ algebra, one finds that $\tau a_z+\bar{\tau} a_{\bar z}$ is proportional to $a_\varphi\,$. Let us define the coefficient of proportionality as $\tilde \tau$,
\be
\tau a_z+\bar{\tau} a_{\bar z}=\tilde{\tau} a_\varphi\, ,\label{spin2taut}
\ee
which we will interpret shortly. It is easy to verify that
\be \tilde{\tau}={\tau-\bar{\tau}\mu_2\over1-\mu_2}\,.\label{ttspin2}
\ee

\noindent This expression can be understood as follows: the deformed metric is conformally equivalent to the standard flat metric $d\tilde{z}d\bar{\tilde{z}}\,$ on the torus, but where the identifications are now $\tilde{z}\simeq \tilde{z} + 2\pi \simeq \tilde{z} + 2\pi \tilde{\tau}$, with $\tilde{\tau}$ given by \eqref{ttspin2}. Smoothness at the horizon is equivalent to the holonomy condition along the thermal circle
\be
\text{Tr}\left[(\tau a_z+\bar{\tau} a_{\bar z})^2\right]=-\frac{1}{2}\,,
\ee
which amounts to
\bea
\tilde \cL = -\frac{k}{4\tilde \tau^2}\,.\label{holosl2}
\eea

Since the metric written in terms of the tilde charges is again that of a usual BTZ black hole we immediately know that the Bekenstein-Hawking entropy, computed with any of the traditional methods, will be
\begin{align}\label{sbh}
S ={}&\frac{Area}{4G_3}=
 2\pi \sqrt{k\, \tstress}+2\pi \sqrt{k\, \tstressbar}\,.
\end{align}
\noindent We then see that $\tilde{\tau}$ is actually the chemical potential conjugate to $\tilde{\cL}$, satisfying
\be
\tilde{\tau}={i\over 2\pi }{\delta S\over \delta\tilde{\cL}}\,.\label{spin 2 tilde tau}
\ee
The partition function is now defined as\footnote{In order to simplify the notation, $\cL$ and $\cW$ (and their barred counterparts) will be understood as operators when written inside traces, and as the expectation values of said operators otherwise.}
\be
Z(\tilde{\tau},\bar{\tilde{\tau}})\equiv \text{Tr}_{\mathcal{H}}\left[ \exp\left( 2\pi i \tilde{\tau}\tilde{\cL}-2\pi i \bar{\tilde{\tau}}\bar{\tilde{\cL}} \right)\right] .
\ee
Modular invariance relates the partition function at high temperature to the degeneracy of the ground state
\be
\ln Z(\tilde{\tau},\bar{\tilde{\tau}})=\ln Z\left(-1/\tilde{\tau},-1/\bar{\tilde{\tau}}\right)\sim -{2\pi i\over \tilde{\tau}}\tilde{\cL}_0+{2\pi i\over \bar{\tilde{\tau}}}\bar{\tilde{\cL}}_0={\pi  i k\over 2 }\left(\frac{1}{\tilde{\tau}} - \frac{1}{\bar{\tilde{\tau}}}\right)
\ee
where the $\sim$ symbol denotes the saddle point approximation, and we used the ground state stress tensor $\tilde \cL_0=\bar{\tilde{\cL}}_0=-k/4\,$. After a Legendre transformation, and using the holonomy condition \eqref{holosl2}, we see that the microscopic entropy in the Cardy regime reproduces the macroscopic Bekenstein-Hawking entropy \eqref{sbh}.
\noindent In other words, in the presence of the deformation, the black hole entropy takes the usual form as predicted by Cardy's asymptotic growth of states in a unitary CFT, when written in terms of the rescaled stress tensor defined in \eqref{new stress}.

For the sake of comparison, one could define a ``holomorphic'' entropy as
\bea
\tau -\bar \tau \mu_2 = \frac{i}{2\pi } \frac{\delta S_{\text{hol}}}{\delta \cL}\,,\qquad -\bar{\tau} + \tau \bar{\mu}_2 = \frac{i}{2\pi } \frac{\delta S_{\text{hol}}}{\delta \bar{\cL}}\,,
\eea
associated with the partition function
\bea
Z_{\text{hol}}(\tau,\alpha,\bar{\tau},\bar{\alpha}) = \text{Tr}_{\mathcal{H}}\left[ e^{2\pi i \left(\tau \cL + \alpha \cL - \bar{\tau}\bar{\cL} - \bar{\alpha}\bar{\cL}\right)}\right] ,
\eea

\noindent where we defined $\alpha = -\bar\tau \mu_2$ and  $\bar{\alpha} = -\tau \bar{\mu}_{2}\,$. While these quantities are perfectly well-defined in terms of a CFT with zero modes $\stress$, $\stressbar$ and action deformed by $\int d^2x \left(\mu_2 \stress-\bar{\mu}_{2}\stressbar\right)$, the resulting entropy
\bea
S_{\text{hol}} = 2\pi \sqrt{k\, \stress} + 2\pi\sqrt{k\,\stressbar}
\eea
does not agree with the standard BTZ entropy \eqref{sbh}. The insight that we gain here is that the bulk-to-boundary dictionary that allows to microscopically reproduce the BTZ entropy involves the tilded zero modes \eqref{new stress} instead of the original zero modes $\cL,\bar\cL\,$. In \cite{Compere:2013aa}, two of the authors argued that the deformations for the higher spin fields behaves qualitatively in the same way, and we will explore this in detail in the next sections.

Let us finally comment on boundary conditions with fixed spacetime dependent deformation parameters  $\mu_2(x^\pm)$, $\bar\mu_2(x^\pm)$. The boundary metric is written in \eqref{bndmetric}, and  we assume that the boundary topology is a cylinder as before.  Applying the uniformization theorem, there exists a Liouville field $\Phi(x^\pm)$ and coordinates $\tilde x^\pm$ such that the metric is explicitly conformally flat
\bea
ds_{(0)}^2 = -l^2 e^{\Phi(x^\pm)} d\tilde x^+ d\tilde x^-\,.
\eea
We then apply the change of coordinates
\begin{align}
x^\pm ={}&
 x^\pm(\tilde x^+,\tilde x^-) + O(e^{-\rho}), \\
\rho ={}&
 \tilde \rho -\frac{1}{2}\ln \Phi (x^\pm)+ O(\rho^{-1}).
\end{align}
Since the leading behavior of the metric is AdS$_3\,$, the Fefferman-Graham theorem holds, and one can choose the subleading terms in $\rho$ in the diffeomorphism such that  the Fefferman-Graham form of the metric is restored. Note that since the integrated conformal anomaly vanishes, radial diffeomorphisms do not modify the on-shell action \cite{deHaro:2000xn,Papadimitriou:2005ii}. We can then use the observation of  \cite{Banados:1998gg} that the most general solution of Einstein's equations with a flat boundary metric is given by \eqref{AAdS3 metric}. We therefore mapped the Dirichlet problem with arbitrary deformations to the standard Brown-Henneaux form (with tilded fields and coordinates).  Following the field redefinition, the two fluctuating functions $\stress,\stressbar$ which obey the Virasoro Ward identities \eqref{VW} can be therefore mapped to the two holomorphic and anti-holomorphic fluctuating functions $ \tstress, \tstressbar$ in terms of tilded coordinates. The discussion of thermodynamics and microscopic counting from a CFT is then similar to the case of constant deformations.

\section{The $sl(3,\mathds{R})$ case}\label{sec:spin3}

The simplest higher spin gauge theory is the Chern-Simons $sl(3,\mathds{R}) \oplus sl(3,\mathds{R})$ theory in the principal embedding. From the $3d$ bulk point of view, this theory contains a spin-2 field (i.e. the metric) non-linearly coupled to a spin-3 field. The spectrum of the dual $(1+1)$-dimensional CFT contains the stress tensor and two currents of weight $(3,0)$ and $(0,3)$, respectively. The asymptotic symmetry algebra in the presence of non-trivial spin-3 deformation parameters $(\mu,\bar{\mu})\,$\footnote{In order to simplify the notation and facilitate comparison with the recent literature, in the present section we adopt the notation $\mu_{3} \to -\mu$, $\bar{\mu}_{3} \to -\bar{\mu}\,$.} was obtained in perturbation theory in \cite{Compere:2013aa} where it was shown to consist of two copies of the $\cW_3$ algebra, exactly as when no deformation is present \cite{Henneaux:2010xg,Campoleoni:2010zq}. One outcome of the analysis of \cite{Compere:2013aa} is the identification of the energy from the zero modes of the $\cW_3$ algebra at finite $\mu$, $\bar{\mu}$ which differs from previous definitions in the literature \cite{Perez:2012cf,Banados:2012ue,Perez:2013xi,deBoer:2013gz}.

In this section, we will first identify a gauge-invariant definition of the $\cW_3$ zero modes obtained with canonical methods. We will then show that the canonical entropy proposed in \cite{Banados:2012ue,Perez:2013xi} and generalized in \cite{deBoer:2013gz} can be given a microscopic interpretation using the formalism of \cite{Gutperle:2011kf,Kraus:2011ds,Gaberdiel:2012yb} together with our definition of zero modes. Since our main interest is in black hole thermodynamics, we will restrict our analysis to stationary and axisymmetric configurations. From now on, all quantities will be independent of the boundary coordinates $x^\pm$, unless otherwise stated.

We will focus our attention on the spin-3 black hole solution constructed in \cite{Gutperle:2011kf,Ammon:2011nk}. Using the basis of $sl(3,\mathds{R})$ generators  introduced in appendix \ref{app:3}, the reduced connections are\footnote{Note that we have rescaled the charges with respect to \cite{Gutperle:2011kf,Ammon:2011nk} as follows: $2\pi \stress_{\mbox{\tiny there}} = \stress_{\mbox{\tiny here}}$ and $2\pi \wcharge_{\mbox{\tiny there}} = \wcharge_{\mbox{\tiny here}}$, and similarly for the barred quantities.}
\begin{align}\label{hw gauge Gutperle Kraus}
a
={}&
\Bigl( L_{1}-\frac{\mathcal{L}}{k}\, L_{-1}-\frac{\mathcal{W}}{4k}\,W_{-2}\Bigr) dx^{+}
\nonumber\\
&
+\mu \biggl(W_{2}+\frac{2 \mathcal{W}}{k}\,L_{-1}+\left( \frac{ \mathcal{L}}{k}\right)^{2}\, W_{-2}-\frac{2 \mathcal{L}}{k}\,W_0 \biggr) dx^{-}\, ,
\\
\bar{a}
={}&
-\Bigl( L_{-1}-\frac{\bar{\mathcal{L}}}{k}\, L_{1}-\frac{\bar{\mathcal{W}}}{4k}\,W_{2}\Bigr) dx^{-}
\nonumber\\
&
-\bar{\mu} \biggl(W_{-2}+\frac{2 \bar{\mathcal{W}}}{k}\,L_{1}+\left( \frac{ \bar{\mathcal{L}}}{k}\right)^{2}\, W_{2}-\frac{2 \bar{\mathcal{L}}}{k}\,W_0 \biggr) dx^{+}\,.
\label{hw gauge Gutperle Kraus 2}
\end{align}

\noindent Here, $k$ is the level of the embedded $sl(2,\mathds R)$ theory, given by $k = \ell/(4G_3)$, and related to the level $k_{cs}$ of the full theory via $k_{cs} = k/(2\text{Tr}\left[L_0 L_0\right])= k/4\,$. The traces here and below are understood to be taken in the fundamental representation, with the conventions of \ref{app:3}. Notice that the BTZ black hole connections are recovered by setting $\wcharge = \wchargebar =\mu=\bar{\mu}=0$. From the holographic perspective, the black hole solution describes the CFT partition function at finite temperature and finite chemical potentials for the higher spin currents \cite{Kraus:2011ds}.

\subsection{Summary of the holomorphic formalism}
The thermodynamics of spin-3 black holes was first studied in the so-called holomorphic formalism \cite{Gutperle:2011kf}. There, one starts in highest-weight gauge for the $a_+$ component of the connection,
\bea\label{holomorphic hw gauge}
a_+ = L_1 + Q
\eea
with $Q$ constrained by the highest-weight condition $[Q,L_{-1}] = 0$. Since the $\mu,\bar\mu$ deformation does not affect $a_+$, see \eqref{hw gauge Gutperle Kraus}, the charges defined from $Q$ are defined independently of $\mu, \bar \mu$ as
\begin{align}
\cL &= \frac{k}{4}\text{Tr}[L_1 Q] = \frac{k}{8} \text{Tr}\left[a_+^2\right] ,\\
\cW &=-\frac{k}{2} \text{Tr} [L_1^2 Q] = -\frac{k}{6} \text{Tr}\left[a_+^3\right]. \label{defLW}
\end{align}
For future reference, we note that in the highest-weight gauge \eqref{holomorphic hw gauge} the sources $(\tau,\alpha)$ conjugate to the holomorphic charges $(\mathcal{L},\mathcal{W})$ can be identified as the lowest weights in the component of the connection along the contractible cycle of the boundary torus upon Wick rotation to Euclidean signature (c.f. \eqref{z periodicities}), namely
\begin{equation}\label{holo lowest weights}
\tau a_z + \bar \tau a_{\bar z}  = \tau L_1 -\alpha W_{2}+\ldots
\end{equation}

\noindent where $\alpha = \bar\tau \mu$ and the dots represent higher-weight terms fixed by the equations of motion. In the case of constant connections we can use the fact $[a_{z},\tau a_z + \bar \tau a_{\bar z}]=0$ by the equations of motion and therefore expand $\tau a_z + \bar \tau a_{\bar z}$ in a basis of $sl(3,\mathds{R})$ elements built from $a_{z}$ as
\begin{equation}
\tau a_z + \bar \tau a_{\bar z} =  \tau a_z -2  \alpha  \left(a_z^{2} - \frac{1}{3}\text{Tr}\left[a_z^2\right]\mathds{1}\right) \label{conjh}.
\end{equation}

The holomorphic higher spin black hole entropy is a sum of contributions from the unbarred and barred connections. The contribution from the $\bar A$ sector is obtained by simply putting bars on all the charges. The chiral half $S_A^{\text{hol}}$ of the entropy  in the holomorphic formalism can be written as
\bea
S^{\text{hol}}_A(\cL,\cW) = 2\pi \sqrt{k \stress}\, \sqrt{1-\frac{3}{4 C}}\,,
\label{SGK}
\eea
where $C$ is defined from $\cW = 4C^{-3/2}(C-1)\cL^{3/2}k^{-1/2}\,$. We see that the holomorphic entropy is expressed in terms of $\cL,\cW$, the zero modes of the $\cW_3$ algebra at $\mu=\bar\mu=0\,$ (with a similar contribution from the barred sector). Using the holographic dictionary between AdS$_3$ and its dual CFT, it was found that  calculations in the holomorphic formalism  can be mapped to perturbative calculations in the undeformed CFT. However, what is unclear in this formalism is whether this proposed entropy is really the higher spin generalization of the Bekenstein-Hawking entropy in some sense  (see however footnote 3). On the contrary, as mentioned in the introduction, canonical approaches lead to a different ``canonical entropy'' \cite{Campoleoni:2012hp,Banados:2012ue,Perez:2013xi,deBoer:2013gz}. In the following, we will present a canonical formalism that will reproduce the canonical entropy and that will also admit a CFT interpretation. Quite remarkably, we will show that the canonical entropy takes the same functional form as the holomorphic entropy \eqref{SGK}, but with all variables replaced via a field redefinition that follows from the redefinition of the CFT zero modes at finite $\mu,\bar\mu$  deformation.

\subsection{Canonical charges and conjugate potentials }
We propose to define the zero modes of the $\cW_3$ algebra $(\tilde \cL,\tilde \cW)$ in terms of the component of the connection along the spatial circle as
\bea
\tilde \cL &=& \frac{k}{8} \text{Tr}\left[a_\varphi^2\right] ,\\
\tilde \cW &=&-\frac{k}{6} \text{Tr}\left[a_\varphi^3\right] . \label{defLWt}
\eea
\noindent Evaluated on the black hole solution \eqref{hw gauge Gutperle Kraus}-\eqref{hw gauge Gutperle Kraus 2}, we obtain
\bea
\tilde \cL &=& \cL +3\mu \cW + \frac{16}{3k}\mu^2 \cL^2, \\
\tilde \cW &=& \cW+\frac{32\cL^2 \mu}{3k}+\frac{16 \cL \cW \mu^2}{k} - \frac{512 \cL^3 \mu^3}{27k^2}+\frac{16 \cW^2 \mu^3}{k}\, .
\eea
This definition agrees with the zero modes obtained from the asymptotic symmetry algebra analysis performed in \cite{Compere:2013aa}, up to possible $\mathcal{O}(\mu^4)$ corrections. Since the definition \eqref{defLWt} is gauge-invariant, however, it is natural to conjecture that the zero modes of $\tilde \cL$ and $\tilde \cW$ as computed in \cite{Compere:2013aa} will not get any corrections to order 4 and higher in the $\mu$ expansion, which would result in a complete agreement.
Note that the horizon holonomy, depending on the eigenvalues of $a_{\varphi}\,$, was computed in appendix C of \cite{Ammon:2011nk} (see also \cite{Banados:2012ue}). Here, we further identify that these holonomies encode the zero modes of the conformal generators.

In the Euclidean formulation of the theory, regularity of the black hole solution is enforced by requiring that the holonomies of the connection around the contractible cycle of the boundary torus are identical to those of the BTZ black hole, which in the present context implies
\bea\label{holonomy conditions}
\text{Tr}\left[h^2\right] = -8\pi^2\,,\qquad \text{Tr}\left[h^3\right]=0\,,
\eea
\noindent where $h=2\pi\left(\tau a_z + \bar \tau a_{\bar z}\right)\,$. In order to solve the holonomy conditions (\ref{holonomy conditions}), it proves convenient to parameterize the $\cW,\,\bar{\cW}$ in terms of $\cL,\bar\cL$ and auxiliary variables $C,\bar{C}$ as follows
\begin{equation}
\wcharge =
 \frac{4(C-1)}{C^{3/2}}\stress\sqrt{\frac{\stress}{k}}\,,\qquad \wchargebar =
- \frac{4(\bar{C}-1)}{\bar{C}^{3/2}}\stressbar\sqrt{\frac{\stressbar}{k}}\,.\label{defCCt}
\end{equation}
Solving \eqref{holonomy conditions} for  $\tau,\bar\tau,\mu,\bar\mu$ then yields
\begin{align}
-\frac{\bar\tau}{\tau}\mu ={}&
 \frac{3\sqrt{C}}{4(2C-3)}\sqrt{\frac{k}{\cL}}\,,&
\tau ={}&
 \frac{i(2C-3)}{4(C-3)\sqrt{1-\frac{3}{4C}}}\sqrt{\frac{k}{\cL}}\,,\label{muh1}
\\
\frac{\tau}{\bar\tau}\bar\mu ={}& \frac{3\sqrt{\bar C}}{4(2\bar C-3)}\sqrt{\frac{k}{\bar\cL}}\,,&
-\bar\tau ={}& \frac{i(2\bar C-3)}{4(\bar C-3)\sqrt{1-\frac{3}{4\bar C}}}\sqrt{\frac{k}{\bar\cL}}\,.
\label{muh}
\end{align}

\noindent In the non-rotating limit $\stressbar =\stress$ and $\bar{C}=C$, so that $\bar\tau = -\tau\,$, $\wcharge= -\wchargebar$, $\bar{\mu} = -\mu$, reproducing the known results. The solution is defined for $3<C,\bar{C} < \infty\,$ where the upper bound is reached at the BTZ point and the lower bound corresponds to the extremal black hole. For the purposes of comparing results with perturbative expansions in $\mu,\bar\mu$, it is useful to also present the perturbative solution of the holonomy conditions:
\bea
\cL &=& -\frac{k}{4\tau^2} + \frac{5 k\mu^2 \bar \tau^2}{3\tau^6} +\mathcal{O}(\mu^4),\nn\\
\cW &=&-\frac{2k\mu \bar \tau}{3\tau^5}+\frac{160 k \mu^3 \bar \tau^3}{27 \tau^9}+\mathcal{O}(\mu^4), \nn\\
\bar \cL &=& -\frac{k}{4\bar \tau^2} + \frac{5 k\bar \mu^2  \tau^2}{3\bar \tau^6} +\mathcal{O}(\bar \mu^4), \label{solhh} \\
\bar \cW &=& -\frac{2k\bar \mu  \tau}{3\bar \tau^5}+\frac{160 k \bar \mu^3  \tau^3}{27 \bar \tau^9}+\mathcal{O}(\bar \mu^4).\nn
\eea

Having identified the canonical charges as described above, we now turn our attention to the definition of sources which are canonically conjugate to these charges. The more straightforward way to define the sources, that we denote as $\tilde\tau,\, \tilde \alpha,\, \bar{\tilde \tau}, \bar{\tilde \alpha}$, is as follows. We can use the gauge freedom to write $a_{\varphi}$ in highest-weight gauge, namely
\begin{equation}
a_{\varphi} = L_1 + \tilde{Q}
\end{equation}
\noindent where the matrix $\tilde{Q}$ is linear in the canonical charges and satisfies $[L_{-1},\tilde{Q}]=0\,$. The sources can then be identified as the lowest weights in the component of the connection $\tau a_{z} + \bar{\tau}a_{\bar{z}}\,$ as
\begin{equation}
\tau a_{z} + \bar{\tau} a_{\bar{z}}= \tilde{\tau} L_{1} - \tilde{\alpha} W_{2} + \ldots\,,\label{def4}
\end{equation}
where the dots represent higher-weight terms that are fixed by the equations of motion.  This definition can be motivated by the fact that in standard gravity coupled to matter, the thermodynamic conjugate variables to the black hole conserved charges are defined from local properties of the geometry and matter fields close to the Killing horizon. In Chern-Simons theory, it is therefore natural to define the conjugate variables to the charges using the connection along the Euclidean thermal circle, which characterizes the properties of the black hole. More fundamentally, the expression \eqref{def4} is consistent with the variational principle with fixed sources. One can indeed find a boundary term that supplements the Euclidean action in a way such that the variation of the sum of its bulk and boundary pieces leads (on-shell) to
\begin{equation}\label{varp}
\left. \delta I^{(E)}\right|_{os} =  -2\pi i \int_{\partial M}\frac{d^{2}z}{4\pi^{2}\text{Im}(\tau)} \Bigl(\tilde \cL \delta \tilde \tau + \tilde \cW \delta \tilde \alpha - \bar{\tilde \cL} \delta \bar{\tilde \tau} -\bar{\tilde \cW} \delta \bar{\tilde \alpha} \Bigr).
\end{equation}
This derivation uniquely fixes the coefficients of the lowest-weight generators in \eqref{def4}. Instead of presenting the details of this derivation for the spin-3 case, we refer the reader to section \ref{bndterm} where the variational principle is discussed in full generality.

In the special case of constant connections, the equations of motion imply
\bea
\left[\tau a_z + \bar \tau a_{\bar z}\, , a_\varphi\right] = 0\,.
\eea
Since $a_\varphi$ contains all the information about the charges, it is natural to expand the component of the connection along the Euclidean thermal circle in a basis of $sl(3,\mathds{R})$ elements built out of $a_\varphi$ as (see \cite{Banados:2012ue} also)
\bea
\tau a_z + \bar \tau a_{\bar z} = \tilde{\tau} a_\varphi -2 \tilde \alpha  \left(a_\varphi^{2}-\frac{1}{3} \text{Tr}\left[a_\varphi^2\right] \mathds{1}\right),\label{conj}
\eea

\noindent where $\mathds{1}$ is the three-dimensional identity matrix, and similarly for the barred sector. The numerical coefficients in the latter expression cannot be fixed from these arguments alone, but can be adjusted so that the definitions \eqref{def4} and \eqref{conj} exactly agree. As opposed to \eqref{def4}, \eqref{conj} does not require to be in highest-weight gauge for $a_{\varphi}\,$, so in particular it can be evaluated in the gauge used in \eqref{hw gauge Gutperle Kraus}, which leads to
\bea
\tilde\tau &=& \frac{\tau - \frac{16 \mu^2}{3k}\Bigl( \left(\tau+2\bar\tau\right)\cL+3\mu \bar \tau W \Bigr)}{1-\frac{16 \mu^2}{k}(\cL+\mu \cW)}\,,\\
\tilde\alpha &=& \frac{\mu(\bar\tau-\tau)}{1-\frac{16 \mu^2}{k}(\cL+\mu \cW)}\,.\label{taut}
\eea
This is the analogue for spin-3 of the expression \eqref{ttspin2} for the spin-2 case. The chiral temperatures are therefore modified in the presence of spin-3 deformations, in parallel with the modification of the conformal zero modes. Note that contrary to the spin-2 case,  the chemical potentials are now field-dependent. In terms of the perturbative $\mu,\bar\mu$ expansion we obtain
\bea
\tilde \tau &=& \tau +\frac{8}{3}\frac{\bar\tau-\tau}{\tau^2}\mu^2 +\mathcal{O}(\mu^4),\\
\tilde \alpha &=& (\bar \tau - \tau)\mu - 4\frac{\bar \tau- \tau}{\tau^2}\mu^3 +\mathcal{O}(\mu^5).\label{solatilde}
\eea

Given that the structure of \eqref{conjh} and \eqref{conj} is formally the same, we can immediately write down the solution to the holonomy conditions in tilded variables as (\ref{defCCt})-(\ref{muh}) with tildes on everything,
\bea
\tilde{\wcharge} &=&
 \frac{4(\tilde{C}-1)}{\tilde{C}^{3/2}}\tilde{\stress}\sqrt{\frac{\tilde{\stress}}{k}}\,,\\
 -{\bar{\tilde{\tau}}\over\tilde{\tau}}\tilde{\mu} &=&{}
 \frac{3\sqrt{\tilde{C}}}{4(2\tilde{C}-3)}\sqrt{\frac{k}{\tilde{\cL}}}\,,\\
\tilde{\tau} &=&{}
 \frac{i(2\tilde{C}-3)}{4(\tilde{C}-3)\sqrt{1-\frac{3}{4\tilde{C}}}}\sqrt{\frac{k}{\tilde{\cL}}}\,,
\eea
and similarly for the barred sector. Using the definition of $\tilde C$, $\tstress$ and $\twcharge$ and substituting $\mu$ using \eqref{muh} we find the additional relation
\be
\tilde{C}={C(2C-3)^2-9{\tau\over\bar \tau}(2C-3)(C-1)+3\left({\tau\over\bar \tau}\right)^2C^2\over(2C-3)^2\left(1-{\tau\over\bar{\tau}}\right)^2}\,.
\ee

\subsection{Canonical entropy}\label{sec:3}
Let us first quickly review Wald's derivation \cite{Wald:1993nt} of the stationary black hole entropy in the metric formalism for general diffeomorphic-invariant theories without matter. The main point in Wald's derivation is the identification of the canonical charge $\mathcal Q_\xi$ associated with the Killing symmetry
\bea
\xi = \beta \left( \frac{\p}{\p t} +\Omega \frac{\p}{\p \varphi}\right)
\eea
as the entropy $S$, i.e.
\bea
S = \mathcal Q_{\xi}\,.
\eea
The Killing symmetry is naturally defined in terms of the properties of the horizon: $\beta$ is the inverse of the Hawking temperature and $\Omega$ is the angular velocity. Defining
\bea
\tau = \frac{i\beta }{2\pi }(1+\Omega)\,,\qquad \bar\tau = \frac{i\beta}{2\pi}(-1+\Omega)\,,
\eea
the Killing generator can be written as
\bea
\xi = -2\pi i \left( \tau l_0 -\bar \tau \bar l_0 \right)
\eea
where $\p_t = l_0+\bar l_0$ and $\p_\varphi = l_0 - \bar l_0\,$. Since the Noether charge is linear in its symmetry generator and conserved, and since $l_0,\bar l_0$ are associated with the $\cL_0,\bar \cL_0$ charges, the first law
\bea
\delta S = -2\pi i \left( \tau \delta \cL_0 - \bar \tau \bar\cL_0 \right)
\eea
holds by construction. Specializing to $sl(2,\mathds R)$ Chern-Simons theory, the gauge symmetry $(\Lambda,\bar\Lambda) = (\xi^\mu A_\mu,\xi^\mu \bar A_\mu)$ associated with the Killing symmetry reads
\bea
(\Lambda,\bar \Lambda) = -2\pi i \left(\tau A_+ - \bar \tau A_-, \tau \bar A_+ - \bar \tau \bar A_-\right).\label{gs}
\eea
We will now show that the same gauge symmetry \eqref{gs} is more generally canonically associated with the black hole entropy in $sl(3,\mathds R)$ Chern-Simons theory.

In the case of $sl(3,\mathds R)$ Chern-Simons theory in the principal embedding, the first law should read
\bea
\delta S = -2\pi i \left(\tilde \tau \delta \tilde \cL + \tilde \alpha \delta \tilde \cW - \text{barred}\right) \label{first1}
\eea
where the barred sector has the same structure (with barred variables). It was shown in \cite{Compere:2013aa} that the infinitesimal canonical Noether charge associated with the constant gauge parameter generators $(\tilde \eps,\tilde \chi)$ in the black hole background \eqref{hw gauge Gutperle Kraus} can be written as
\bea
\delta \cQ_{(\tilde \eps,\tilde \chi)}= \tilde \eps \delta \tilde \cL - \tilde \chi \delta \tilde \cW\,.
\eea
One has therefore to prove that the Noether charge associated with
\bea
(\tilde \eps,\tilde \chi) = -2\pi i (\tilde \tau, - \tilde \alpha)
\eea
is integrable (i.e. a $\delta$-exact quantity); since the result (after summing up the unbarred and barred sectors) would obey the first law \eqref{first1}, it would be identified as the black hole entropy.

The relationship between the gauge transformation $\delta A_\mu = \p_\mu \Lambda$ and the gauge parameter generators $(\tilde \eps,\tilde \chi)$ was worked out in \cite{Compere:2013aa} up to $O(\mu^4)$ in perturbation theory. The gauge transformation parameter $\Lambda = e^{-\rho L_0} \lambda e^{\rho L_0}$ is given by
\bea
\lambda = \eps L_1 +\chi W_2 + \cdots\label{lambda4}
\eea
where the dots denote higher-weight $sl(3,\mathds R)$ generators and $(\eps,\chi)$ are given by
\bea
\left(\begin{array}{c} \eps \\ \chi \end{array}\right) = \left(\begin{array}{cc} 1& -\frac{32\cL}{3k}\mu - \frac{16\mu^2}{k}\cW \\ -\mu & 1-\frac{16 \cL}{3k}\mu^2 \end{array}\right)
\left(\begin{array}{c} \tilde \eps \\ \tilde \chi \end{array}\right)+O(\mu^4)\,.
\eea
This non-trivial linear transformation arises from the requirement that the conserved charges $\tilde \cL$, $\tilde \cW$ are the zero modes of the $\cW_3$ algebra up to $O(\mu^4)$ corrections. After using the holonomy conditions \eqref{solhh} we find
\bea
\left(\begin{array}{c} \eps \\ \chi \end{array}\right) = \left(\begin{array}{c} -2\pi i \tau \\ 2\pi i \mu \bar\tau \end{array}\right) + O(\mu^4)\,.
\eea
We now recognize that the lowest $sl(3,\mathds R)$ weights in the connection $\tau a_+ - \bar\tau a_-$ match with \eqref{lambda4} up to $O(\mu^4)$. Therefore, we showed that the gauge symmetry that gives the right-hand side of the first law \eqref{first1} is generated by
\bea
\lambda = -2\pi i \left(\tau a_+ - \bar\tau a_- \right),
\eea
up to $O(\mu^4)$ corrections. This generator has a very natural interpretation in Euclidean signature: it is $\lambda = -i h\,$, where $h=2\pi (\tau a_z+\bar\tau a_{\bar z}) = \oint a$ is the integral of the reduced connection around the thermal circle. It is therefore natural to conjecture that the $O(\mu^4)$ corrections will vanish.

It remains to prove that the left-hand side of the first law \eqref{first1} is integrable and derive the actual formula for the entropy. Restoring the barred sector, the infinitesimal canonical charge integrated over the $\varphi$ circle is given by \cite{Regge:1974zd}
\begin{equation}
\sdelta \cQ_{\lambda =- ih,\bar\lambda =-i\bar h}  = -  \frac{i k_{cs}}{2\pi}  \int_0^{2\pi} d\varphi\, \text{Tr}\left[\delta a_\varphi\, h - \delta \bar a_\varphi\, \bar{h}\right],
\end{equation}
where the notation $\sdelta$ emphasizes that we do not yet know whether the quantity is $\delta$-exact. Since  $[a_\varphi,h]=0$ from the equations of motion and since smoothness of the solution implies that there exists a group element $u$ such that $h = u^{-1}\left(2\pi i L_0\right) u\,$, one has \cite{deBoer:2013gz}
\bea
\text{Tr}\left[a_\varphi \delta h\right] = \text{Tr}\left[\bar a_\varphi \delta \bar{h}\right]  = 0\,.
\eea
Therefore, the charge is integrable and its integral $\sdelta \cQ_\Lambda = \delta S$ is given by the entropy
\bea
S = - i k_{cs} \text{Tr}\left[ a_\varphi h -\bar a_\varphi \bar{h}\right],\label{SdeBJ}
\eea
where $h = 2\pi\left(\tau a_z + \bar \tau a_{\bar z}\right)$, $\bar{h} = 2\pi\left(\tau \bar{a}_z + \bar{\tau} \bar{a}_{\bar z}\right)$. The result exactly matches with the canonical entropy derived in \cite{deBoer:2013gz}.

\subsection{Matching to the (deformed) CFT}\label{subsec:matching}

According to the asymptotic symmetry algebra analysis \cite{Compere:2013aa}, the deformed dual theory is still governed by $\cW_3 \oplus \cW_3$ symmetry with central charge $c=6k\,$. The zero modes of symmetry generators are given by  $\tstress,\twcharge,\bar \tstress,\bar \twcharge$.  This suggests the following dictionary from the bulk variables to CFT variables
\begin{alignat}{3}
 \tilde{\cL}&= \cL_{0;\text{CFT}} \,,& &\qquad & \twcharge &= \cW_{0;\text{CFT}}\,, \\
\tilde{\tau}&=\tau_{\text{CFT}}\,,& &\qquad & \tilde{\alpha}&=\alpha_{\text{CFT}}\,.
\end{alignat}

\noindent It is then natural to rewrite the canonical entropy in terms of the tilded variables, and see whether it suggests any interpretation in the deformed CFT.
It turns out that by using the relations between the tildes variables and untilded variables, the canonical entropy in \eqref{SdeBJ} takes exactly the same functional form as the holomorphic entropy of \cite{Gutperle:2011kf}, but as a function of the tilded charges:
\bea
S= S_A+S_{\bar A} = 2\pi \sqrt{k \tstress}\, \sqrt{1-\frac{3}{4 \tilde{C}}} +2\pi \sqrt{k \bar{\tstress}}\, \sqrt{1-\frac{3}{4 \bar{\tilde{C}}}}\,.
\eea
This formula can be more easily proven by using as intermediate expressions
\bea
S_A &=& 2\pi\sqrt{k\cL}\left(1-{3\over2C}\right)^{-1}\sqrt{1-{3\over4C}}\left(1-{3\over2C}\left(1+{{\tau}\over\bar\tau}\right)\right) \\
&=& -2 \pi i \left(2\tau \cL +\frac{32}{3k}\bar\tau \mu^2 \cL^2 + 3(\tau+\bar\tau)\mu\cW\right).
\eea

The free energy, which is as usual related to the entropy by a Legendre transformation, can be written as
 \bea \ln Z_{bh}=-2\pi i\Big(\tilde{\tau}\tilde{\cL}+\tilde{\alpha}\twcharge\Big).
\eea
It should then correspond to the logarithm of the CFT partition function
\be
\ln Z_{\text{CFT}}=\ln \text{Tr}_{\mathcal{H}}\left[
 e^{2\pi i\left(\tau_{\text{CFT}}\hat{\cL}_{0;\text{CFT}}+\alpha_{\text{CFT}}\hat{\mathcal{W}}_{0;\text{CFT}}\right)} \right].
\ee
For $sl(3,\mathds R)$, a direct CFT calculation is not available. However, as we will see explicitly in the next section, our formalism matches exactly with a direct CFT calculation for spin-3 black holes in the theory based on the $\text{hs}[\lambda]$ algebra.

\subsection{AdS$_{3}$ vacua in the presence of higher spin deformations}
A natural question that arises when considering $\mu,\bar\mu$ deformations is the existence of a vacuum. Is there a natural definition of an $sl(2,\mathds{R}) \oplus sl(2,\mathds{R})$ invariant vacuum for each value of $\mu,\bar \mu$? From the analysis of \cite{Compere:2013aa}, the asymptotic symmetry algebra was shown to be isomorphic to the undeformed asymptotic symmetry algebra in perturbation theory in $\mu,\bar\mu$. It shows that the global AdS$_3$ vacuum can still be defined by fixing the holonomies around the $\varphi$ circle to be trivial, and equal to their values for the undeformed AdS$_3$ vacuum.  A first hint that the symmetry algebra is preserved at finite $\mu,\bar\mu$ is that the zero modes of the asymptotic symmetry algebra are defined in a gauge-invariant way for any finite $\mu,\bar\mu$ as shown in \eqref{defLWt}. We will propose here a gauge-invariant definition of vacua at finite $\mu,\bar\mu$. As a non-trivial check, we will show  from first principles that these vacua have the same number of symmetries as the undeformed AdS$_3$ vacuum. This suggests that the symmetry-preserving perturbative expansion has a finite radius of convergence, providing evidence, at finite $\mu\,\bar\mu$, of the picture advocated in  \cite{Compere:2013aa}.

We propose the definition of vacua at finite values of $\mu,\bar\mu$ as the connections which have a trivial holonomy around the $\varphi$ circle, so that the eigenvalues of $a_\varphi$ match the ones of the global AdS$_3$ solution. This amounts to set the values of the zero modes to the values of the undeformed global AdS$_3$ vacuum,
\begin{equation}
\tilde \cL  =\bar{\tilde \cL} =  -\frac{k}{4}\,,\qquad \tilde \cW = \bar{\tilde \cW} =0\,.
\end{equation}

\begin{figure}[htb]
\begin{minipage}{0.45\textwidth}
\begin{flushleft}
\resizebox{\textwidth}{!}{\mbox{\includegraphics{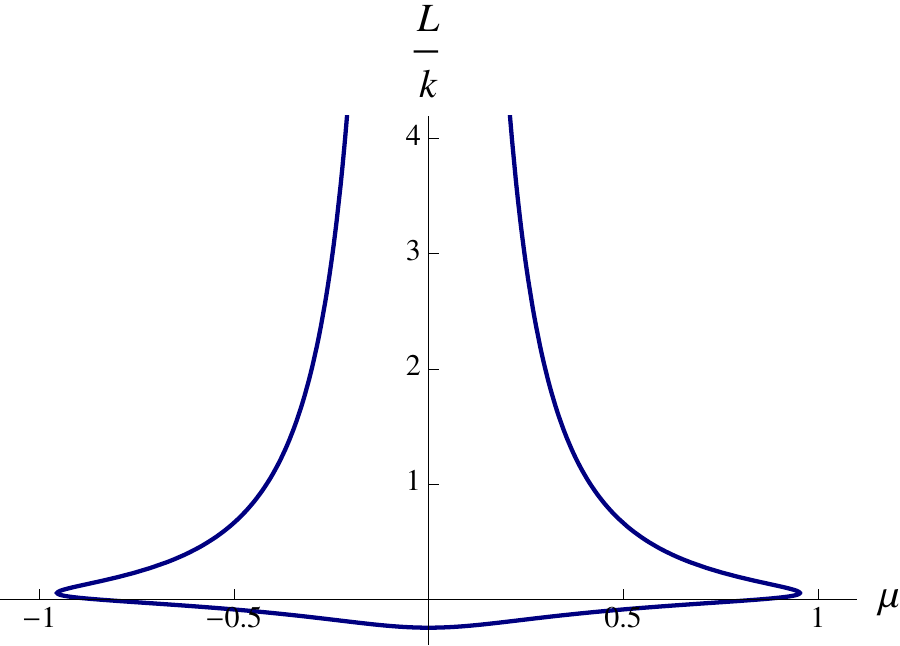}}}
\caption{Plot of $\cL$ as a function of the deformation parameter $\mu$ for the vacuum solutions.}
\label{fig:vacuumL}
\end{flushleft}
\end{minipage}
\begin{minipage}{0.10\textwidth}\mbox{}$\qquad$
\end{minipage}
\begin{minipage}{0.45\textwidth}
\begin{flushright}
\resizebox{\textwidth}{!}{\mbox{\includegraphics{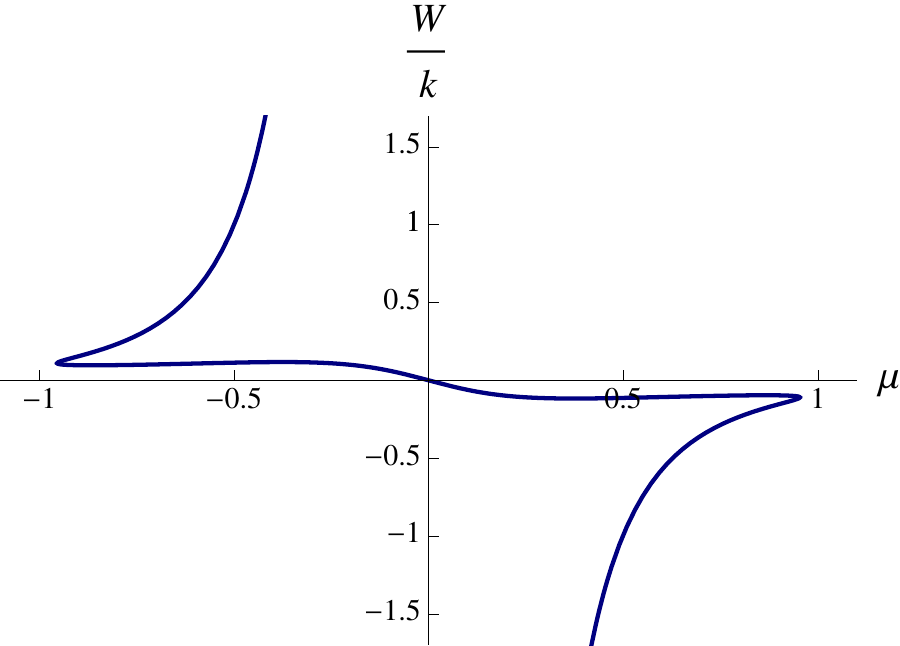}}}
\caption{Plot of $\cW$ as a function of the deformation parameter $\mu$ for the vacuum solutions.}
\label{fig:vacuumW}
\end{flushright}
\end{minipage}
\end{figure}

Let us consider the unbarred sector.  Using the definition of the zero modes in terms of the holonomies we can solve for $\cW$ in terms of $\cL\,$,
\bea
\cW = -\frac{1}{3\mu}\left(\cL +\frac{k}{4}\right)-\frac{16\mu}{9k}\cL^2\,, \label{eqW1}
\eea
and one is left with the following quartic equation for $\cL$ (and similarly for $\bar\cL$),
\begin{align}
0={}&
\left(36 \mu ^2-27\right)-\left(144 \mu ^2+108\right) \cL+\left(1536 \mu ^4+1728 \mu ^2\right)  \cL^2
\nonumber\\
&
-9216 \mu ^4 \cL^3 +16384 \mu ^6 \cL^4 \,.
\label{eqL1}
\end{align}
Note that the equations \eqref{eqW1}-\eqref{eqL1} can also be formally obtained by writing the holonomy conditions for the spin-3 black holes with $\tau=\bar\tau = 1\,$. It turns out that there are two real branches of solutions around $\mu = 0\,$. One branch starts at the undeformed AdS$_3$ vacuum value $\cL=-\frac{k}{4}\,$ while the other starts at $\cL=\infty\,$. The two branches merge at the critical values for the deformation parameter
\bea
\mu_c = \pm \frac{3}{8}\sqrt{3+2\sqrt{3}}\approx \pm 0.95\,.
\eea
There is no real solution to the holonomy conditions for $|\mu| > |\mu_c|\,$. The values for $\cL,\cW$ are plotted as a function of $\mu$ in figures \ref{fig:vacuumL}-\ref{fig:vacuumW}. Interestingly, the bound $|\mu | \leq \mu_c$ is close to the bound $|\mu_2 | < 1$ for the spin-2 case. In conclusion, for all values $-\mu_c \leq \mu \leq \mu_c$ there are two independent vacua which coincide at $\pm\mu_c\,$.

Gauge symmetries, also called reducibility parameters, are defined as gauge generators which leave the fields invariant. The vacuum solution which asymptotes to the ordinary AdS$_3$ vacuum at $\mu = 0$ is $sl(2,\mathds{R}) \oplus sl(2,\mathds{R})$ invariant around $\mu = 0$ as a consequence of the existence of a $\cW_3 \oplus \cW_3$ asymptotic symmetry algebra in the perturbative $\mu$, $\bar\mu$ expansion \cite{Compere:2013aa}. Indeed, using the conventions of \cite{Compere:2013aa} and looking at the unbarred sector, the conditions $\delta_{\eps_0,\chi_0} \tilde \cL =0\,$, $\delta_{\eps_0,\chi_0} \tilde \cW = 0$ have three linearly independent solutions for the gauge parameters $(\eps_0,\chi_0)=(1,0),(e^{\pm ix^+},0)\,$ which generate the $sl(2,\mathds{R})$-left algebra. In $sl(3,\mathds{R})$ Chern-Simons theory, the vacuum AdS$_3$ has five additional symmetries $(\eps_0,\chi_0)=(0,1)\,$, $(0,e^{\pm i x^+})\,$, $(0,e^{\pm 2i x^+})$ which correspond to symmetries associated with the spin-3 field. Here, we will investigate if the number of symmetries (8 in the left sector) is preserved upon turning on the $\mu$ deformation.

Symmetry generators obey the following equations
\begin{align}
0={}&
 \delta_{\eps,\chi}\cL =-2\cL \p_+ \eps +\frac{k}{2}\p_+^3 \eps +3\cW \p_+ \chi \,,
 \\
0={}&
\delta_{\eps,\chi}\cW = -3 \cW \p_+ \eps -\frac{10}{3}\cL \p_+^3 \chi  +\frac{k}{6}\p_+^5 \chi +\frac{32}{3k}\cL^2 \p_+ \chi\,,
\end{align}
for $\cL,\cW$ solutions to \eqref{eqW1}-\eqref{eqL1}. Moreover, the gauge generators are restricted to obey the boundary conditions, which amounts to
\begin{equation}
\p_- \chi = 2\mu\, \p_+ \eps\,,\qquad \p_- \eps = -\frac{2\mu}{3}\p_+^3 \chi +\frac{32\mu}{3k}\cL\, \p_+ \chi\,.\label{eq:w}
\end{equation}
Let us solve these equations. Since $\cL$ is constant, the latter system is a linear wave equation with a non-linear dispersion relation. Imposing periodicity along the $\varphi$ coordinate, we find that the most general solution to \eqref{eq:w} at finite $\mu$ (besides $\eps$ or $\chi$ constant) is labeled by the wave number $m \in \mathds{Z}_0$ along $\varphi$ and the momentum $k_+$ along $x^+$ as
\bea
(\eps,\chi) = \left(k_+-m \,, \,2\mu k_+\right)e^{i(k_+-m) x^- +i k_+ x^+ }\label{ec}
\eea
where the dispersion relation can be written as a quartic equation for $k_+$,
\bea
\left(k_+- m\right)^2 = \frac{4\mu^2}{3}k_+^2 \left(k_+^2 +\frac{16\cL}{k}\right).\label{disp}
\eea
Note that this general solution at $\mu \neq 0$ reproduces only 6 out of the 8 reducibility parameters in the limit $\mu \rightarrow 0\,$, since it fixes the relative coefficient between $\eps$ and $\chi$ when $m=\pm 1$. The remaining two reducibility parameters for $m = \pm 1$ can be obtained by taking the limit $\mu \rightarrow 0$ in the equations first. Nevertheless, we will see below that there will be 8 independent solutions at finite $\mu\,$, which degenerate to 6 solutions in the $\mu \rightarrow 0$ limit.

Plugging the solution \eqref{ec} into the symmetry equations we obtain
\begin{align}
12 k_+ \frac{\cW}{k}\mu &= (k_+-m)\left(k_+^2 +4\frac{\cL}{k}\right),
\\
9\left(k_+ - m\right)\frac{\cW}{k} &= k_+ \left(k_+^2+\frac{4\cL}{k}\right)\left(k_+^2 +\frac{16 \cL}{k}\right)\mu\,.
\label{eq45}
\end{align}
After some algebra we find that, given \eqref{eqW1}-\eqref{eqL1}, a solution to \eqref{disp}-\eqref{eq45} exists only for $m=\pm 1$ or $m =\pm 2\,$. For $m = \pm 1\,$, the solution can be constructed after solving
\begin{equation}
9 \mp 72 k_++198 k_+^2 \mp 216 k_+^3+k_+^4 \left(168 \mu ^2+81\right) \mp 96
   k_+^5 \mu ^2 +72 k_+^6 \mu ^2+16 k_+^8 \mu ^4=0\,.
\end{equation}
For $m = \pm 2\,$, one has to solve instead
\bea
12 \mp 24 k_++9 k_+^2+4 k_+^4 \mu ^2 = 0\,.
\eea
Moreover, one has to choose the branches for $k_+$ such that $\cL\,$ given by
\bea
\frac{\cL}{k} = -\frac{k_+^2}{16}+\frac{3\left(k_+-m\right)^2}{64 k_+^2 \mu^2}
\eea
is real. For $m=1$ and $|\mu| < |\mu_c|$, we find two pairs of complex conjugate momenta $k_+$ that obey the requirements. These two pairs reproduce the two branches of $\cL$ of figure \ref{fig:vacuumL}. The imaginary part of these complex roots degenerate to 0 in the limit $\mu \rightarrow 0\,$. For $m = 2\,$, we obtain two real solutions for $k_+\,$. These two solutions also reproduce the two branches of $\cL$ of figure \ref{fig:vacuumL}. The analysis is similar for $m =-1,-2$ since only the sign of $k_+$ flips. In conclusion, on the lowest branch of figure \ref{fig:vacuumL} (connected to the undeformed AdS$_3$), we found 8 reducibility parameters: the 2 constants $(\eps,\chi)$, the complex conjugate pair for $m=1$ and the one for $m=-1\,$, and the real solution for $m=2$ and for $m=-2\,$. Following the same logic, there are also 8 reducibility parameters on the upper  branch of figure \ref{fig:vacuumL}. Therefore, the number of gauge symmetries is preserved upon turning a finite value of $|\mu| \leq | \mu_c|\,$. The vacua are maximally symmetric, and therefore gauge-equivalent to global AdS$_3\,$.

\section{Generalization to $sl(N,\mathds{R})$ and $\mbox{hs}[\lambda]$}\label{sec:generalization}

Building on the spin-$2$ and spin-$3$ examples that we examined in detail in the previous sections, we will now extend these considerations to arbitrary $N$ and more generally to the theory based on the infinite-dimensional hs$[\lambda]$ algebra. Moreover, we will build a variational principle that is appropriate to the canonical definition of charges and chemical potentials we have provided. We will write the entropy as a bulk Noether charge, and show that it  can be obtained as the Legendre transform of the free energy under these boundary conditions. As a first step, we will review and slightly reformulate the holomorphic formalism, which will help to contrast our respective definitions. Our approach here will be based on the treatment in \cite{deBoer:2013gz}, which extended and generalized earlier work in \cite{Banados:2012ue}. For concreteness, when discussing the $sl(N,\mathds{R})\oplus sl(N,\mathds{R})$ theory we will focus on the principal $sl(2,\mathds{R})$ embedding. As it will be clear from the manipulations below, when written in terms of the connection components our final results will apply to other embeddings as well.

In the Euclidean formulation of the theory we consider the boundary of the bulk manifold $M$ to be topologically a torus with modular parameter $\tau\,$. The Lorentzian light-cone coordinates are analytically continued as $x^{+} \to z$ and $x^{-} \to -\bar{z}\,$, with the identifications
\begin{equation}
z \simeq z+ 2\pi \simeq z +2\pi \tau\,.
\end{equation}
\noindent The volume of the boundary torus is then $\text{Vol}(\partial M) = 4\pi^{2}\text{Im}(\tau)\,$. We define
\be
k_{cs} = \frac{k}{2 \text{Tr}[L_0L_0]}\,,
\ee
such that the Virasoro algebra acting on the standard Brown-Henneaux boundary conditions for the principal $sl(2,\mathds{R})$ embedding has central charge $c=6k\,$.

\subsection{Summary of the holomorphic formalism}\label{subsec:holo}
In this subsection we review and reformulate how thermodynamics works in the holomorphic formalism.  We will generically denote the spin-$j$ charge in the holomorphic formalism by $Q_{j}$ (and similarly for the barred sector), so $(Q_{2},\bar{Q}_{2})$ corresponds to the stress tensor zero modes at zero deformation parameters $(\mu_j,\bar\mu_j)$, $(Q_{3},\bar{Q}_{3})$ to the zero modes of the spin-3 current, and so forth. The treatment of both chiral sectors is entirely analogous, so below we will focus on the unbarred sector for simplicity. The traces in the $\text{hs}[\lambda]$ algebra are understood to be evaluated according to the conventions in appendix~\ref{appendix:hs-lambda}. When $\lambda=N$, these reduce to matrix traces in the fundamental representation of $sl(N,\mathds{R})\,$. Elements in $\text{hs}[\lambda]\oplus\mathds{C}$ are understood to be multiplied using the $\star$ product defined in appendix \ref{appendix:hs-lambda}, but we omit the explicit $\star$ symbol in order to simplify the notation.

In the holomorphic formulation, the charge $Q_j$ can be read off from the $a_z$  component of the connection. Note that $a_z$ is usually written in the highest-weight gauge
\be
a_z=L_1+Q
\ee

\noindent with
\be
Q = \frac{1}{k_{cs}} \sum_{j=2}^\infty \frac{Q_j}{N_{j-1}^j (\lambda)} V_{-(j-1)}^j\,, \label{matrixQ}
\ee

\noindent where we used the normalization $N^j_{j-1}(\lambda)$ defined in \eqref{Ns}. The latter expression is similar to the one used in \cite{Castro:2011iw}, for example. From the highest-weight form for $Q$ we easily obtain
\bea
Q_j = k_{cs} \text{Tr}\left[V_{j-1}^j Q\right] = k_{cs}  (4q)^{j-2}\text{Tr}\left[(L_{1})^{j-1} Q\right],
\eea

\noindent where the second equality follows from \eqref{slN hw generators}. The first few terms in the expansion of $Q$ read\footnote{We note that, when written in terms of $k$ (rather than $k_{cs}$), the expression for $Q$ is independent of the normalization of the trace ($\gamma$) chosen in \eqref{deftr}.}
\begin{equation}
Q = -\frac{Q_{2}}{k}L_{-1} + \frac{5Q_{3}}{16kq^{2}\left(\lambda^{2}-4\right)}W_{-2} -\frac{35Q_{4}}{384 k q^{4}\left(\lambda^{4} - 13\lambda^{2} + 36\right)}J_{-3} + \ldots
\end{equation}

\noindent In the $\lambda=N=3$ example, our conventions for the $sl(3,\mathds R)$ generators in appendix \ref{app:3} correspond to $q=1/2$, $\fudge=1$, and we find
\begin{equation}
Q = -\frac{Q_{2}}{k}L_{-1} + \frac{Q_{3}}{4 k}W_{-2} = \frac{2}{k}\left(\begin{array}{ccc}
0&Q_2 & Q_3 \\
0&0&Q_2\\ 0&0&0
\end{array} \right)
 \,,
\end{equation}
where
 $Q_{2}$ and $Q_{3}$ have simple expressions in terms of traces of $a_z\,$, i.e.
\begin{equation}
Q_{2} =\frac{k}{4} \text{Tr}[L_{1} Q]= \frac{k}{8} \text{Tr}\left[a_{z}^{2}\right],\qquad Q_{3} =\frac{k}{2}\text{Tr}[L_1^2 Q]= \frac{k}{6}\text{Tr}\left[a_{z}^{3}\right].\label{Q23}
\end{equation}
These definitions match the charges in the holomorphic formalism $Q_2 = \cL\,$, $Q_3 = - \cW$ up to an irrelevant sign for the spin-3 charge.

Starting at $j=4$, the expression for $Q_{j}$ involves more terms than just $(a_z)^{j}\,$. After some algebra one obtains, for example,
\begin{align}
Q_4 ={}&
 k_{cs}(4q)^{2}\, \text{Tr}\left[L_1^3 Q\right]
 \nonumber\\
 ={}&
  \frac{k_{cs}}{4}(4q)^{2}\left(\text{Tr}\left[a_z^4\right] - \frac{3}{5\fudge}\frac{3\lambda^{2}-7}{\lambda\left(\lambda^{2}-1\right)} \left( \text{Tr}\left[a_z^2\right]\right)^2 \right).
\end{align}

We would now like to write down the sources that are conjugate to the charges defined above. To this end, we first notice that one can also express the charge $Q_j$ in terms of $a_z$ as
\bea
Q_j= k_{cs} \frac{(4q)^{j-2}}{j} \text{Tr}\left[a_z b_{j-1}(a_z)\right], \label{charges from traces ho}
\eea

\noindent where $b_{j-1}(a_z) $ is a traceless polynomial of degree $j-1$ in $a_z\,$, i.e. satisfying
\be
{d\over d \lambda} b_{j-1}(\lambda a_z)|_{\lambda=1}=(j-1) b_{j-1}(a_z)\,,\qquad \text{Tr}\left[b_{j-1}\right]=0\,.
\ee

\noindent More explicitly, we can write
\bea
b_{j-1}(a_z) = a_z^{j-1} +\sum_{m=2}^{j-2} c_m[a_z] a_z^{j-1-m}+c_{j-1}[a_z] V^{1}_{0}\,,
\eea

\noindent where the coefficients $c_m[a_z]$ are in general products of traces of powers of $a_z$, of total degree $m$, which are fixed by requiring \eqref{charges from traces ho} to be satisfied. The last term is proportional to the identity element $V^{1}_{0}$ and its coefficient is adjusted to make $b_{j-1}$ traceless. Let us write down some of these polynomials explicitly. From the above relations for $Q_{2}$, $Q_{3}$ and $Q_{4}$ it immediately follows that
\begin{align}
b_{1}(a_{z}) ={}&
 a_z\,,
 \\
b_{2}(a_z) ={}&
 a_{z}^{2} - \text{Tr}\left[ a_{z}^{2} \right]\left(\frac{4q}{\fudge \lambda}V^{1}_{0}\right),
 \\
 b_{3}(a_z) ={}&
 a_{z}^{3} - \frac{3}{5\fudge}\frac{3\lambda^{2}-7}{\lambda\left(\lambda^{2}-1\right)} \left(\text{Tr}\left[ a_{z}^{2} \right]\right)^2 a_{z}  -\text{Tr}\left[ a_{z}^{3} \right]\left(\frac{4q}{\fudge \lambda}V^{1}_{0}\right).
\end{align}

Our next observation is that the set $\{b_{j-1} \}_{j \geq 2}$ form a complete basis to the solutions of the equations of motion for constant connections:
\be
\left[a_z, \tau a_{z} +\bar{\tau}a_{\bar{z}}\right]=0\,.
\ee

\noindent Hence, we can expand the component of the connection along the contractible cycle using these polynomials. In the holomorphic formalism, we define the sources $\alpha_j$ (with $j \geq 2$) conjugate to the charges $Q_{j}$ as the coefficients in this expansion, i.e.
\be
\left(\tau a_{z} +\bar{\tau}a_{\bar{z}}\right) = \sum_{j \geq 2} (4q)^{j-2} \alpha_j b_{j-1}(a_{z})\,.
\ee
\noindent It follows that
\be
 \text{Tr}\left[a_z\left(\tau a_{z} +\bar{\tau}a_{\bar{z}}\right)\right] ={k_{cs}^{-1}}
 \sum_{j \geq 2}j\alpha_{j} Q_{j}\,.
\ee
Next, using the definition of $b_{j-1}$ we have
\begin{align}
\text{Tr}\left[L_1 b_{j-1}\right] ={}&
 \text{Tr}\left[L_1 (L_1 + Q)^{j-1} + \dots\right]
 \nonumber\\
 ={}&
  \text{Tr}\left[L_1^j + (j-1) \left(L_1\right)^{j-1}Q\right],
\end{align}
\noindent where the dots in the first line denote terms which are quadratic and higher in the charges $Q_k\,$, whose contribution to the trace is canceled by similar terms coming from the $L_{1}(a_{z})^{j-1}$ term (by definition of $b_{j-1}$). Since $\text{Tr}[L_1^j]=0\,$, we conclude
\begin{equation}
\text{Tr}[L_1 b_{j-1}(a_{z})] =(j-1)\text{Tr}\left[ \left(L_1\right)^{j-1}Q\right].
\end{equation}

\noindent Together with \eqref{charges from traces ho} the latter equation implies
\begin{align}
 \text{Tr}\left[L_{1}\left(\tau a_{z} +\bar{\tau}a_{\bar{z}}\right)\right] ={}&{k_{cs}^{-1}} \sum_{j\geq 2}(j-1)\alpha_{j} Q_{j}\,,
 \label{definition sources 2h} \\
\text{Tr}\left[Q\left(\tau a_{z} +\bar{\tau}a_{\bar{z}}\right)\right] ={}&
{k_{cs}^{-1}} \sum_{j\geq 2}\alpha_{j} Q_{j}\,.\label{definition sourcesh}
\end{align}
\noindent  When restricting to the $sl(N,\mathds{R})$ theory, similar expressions apply for non-principal embeddings, where the sums now run over the appropriate spectrum, and $j$ is replaced by the conformal weight of the corresponding operators.

Going back to the $N=3$ example (where $q=1/2$, $\gamma=1$), the component of the connection along the contractible cycle reads
\begin{equation}
\tau a_{z} +\bar{\tau}a_{\bar{z}} = \left(\begin{array}{ccc}
-\frac{4\alpha_{3}Q_{2}}{3k} & \frac{2\alpha_{2}Q_{2}}{k} + \frac{4Q_{3}\alpha_{3}}{k} & \frac{2\alpha_{2}Q_{3}}{k}  + \frac{8\alpha_{3}Q_{2}^{2}}{k^2}\\
\alpha_{2} & \frac{8\alpha_{3}Q_{2}}{3k} &  \frac{2\alpha_{2}Q_{2}}{k} + \frac{4Q_{3}\alpha_{3}}{k} \\
2\alpha_{3} & \alpha_{2} & -\frac{4\alpha_{3}Q_{2}}{3k}
\end{array} \right).
\end{equation}

\noindent It is worth emphasizing that this is precisely the structure of the spin-3 black hole solution \eqref{hw gauge Gutperle Kraus}-\eqref{hw gauge Gutperle Kraus 2}, provided we identify
\begin{equation}
\alpha_{2} = \tau \,,\qquad \alpha_{3} = -\bar{\tau}\mu\,.
\end{equation}

\noindent Since $\tau$ couples to the stress tensor and $\bar{\tau}\mu$ to the $(3,0)$ current, we recognize the spin-2 and spin-3 sources as defined in \cite{Gutperle:2011kf} (the extra minus sign in $\alpha_{3}$ is related to the fact that $Q_{3} = -\wcharge$ as was pointed out before).

As usual, in a thermodynamic equilibrium configuration the sources (chemical potentials) and expectation values (charges) form conjugate pairs whose values are related in a way compatible with the laws of thermodynamics. In the context of black hole thermodynamics, these relations are obtained by applying smoothness conditions on the solution. For higher spin black holes, the smoothness conditions can be formulated as the requirement that the gauge connections have trivial holonomy around the contractible cycle of the torus \cite{Gutperle:2011kf,Kraus:2011ds}, which in particular implies
\bea
\text{Tr}\left[h^n\right]=\text{Tr}\left[h^n_{\text{BTZ}}\right], \quad n=2,3,\dots
\label{holoeqho}
\eea
\noindent where $h =2\pi\left(\tau a_{z} +\bar{\tau}a_{\bar{z}}\right)$ as before, and $h_{\text{BTZ}}$ denotes the holonomy matrix evaluated in the BTZ solution (this is, with all the higher spin charges and sources turned off). The solution to these equations determines the charges as a function of the sources (or viceversa), so we have
\bea
Q_j=f_j\left(\left\{\alpha\right\}\right), \qquad j\geq 2\,,
\label{holoho}
\eea
\noindent where $\left\{\alpha\right\}$ denotes the set of sources that are switched on in the solution. Given the structure of the $\text{hs}[\lambda]$ algebra, for non-integer $\lambda$ the holonomy conditions demand that the solution carry all the higher spin charges even if they are not explicitly sourced (this is, a charge $Q_{j}$ is generated even if is not supported by the corresponding source $\alpha_{j}$). This is to be contrasted with the $sl(N,\mathds{R})$ theory, where a solution in the BTZ branch carries a higher spin charge only if the corresponding source is turned on. The perturbative expansion of the solution to the holonomy conditions for the $\text{hs}[\lambda]$ black hole with a source for spin-3 charge can be found in \cite{Kraus:2011ds}.

It was shown in \cite{deBoer:2013gz} that the holomorphic free energy and entropy can be written entirely in terms of the connection as\footnote{When restricting to the $sl(N,\mathds{R})$ theory, the generator $L_{1}$ depends on the choice of $sl(2,\mathds{R})$ embedding under consideration.}
\begin{align}
-\beta F_{\text{hol}}  ={}&
   -2\pi i k_{cs}\,\text{Tr}\biggl[\tau\left(\frac{a_{z}^{2}}{2} \right) -\bar{\tau}\left( \frac{\bar{a}_{\bar{z}}^{2}}{2}\right)
     +\left(\bar{\tau} L_{1} a_{\bar{z}}-\tau L_{-1} \,\bar{a}_{z}\right)\biggr]
     \\
S_{\text{hol}} ={}&
-2\pi i k_{cs}\,\text{Tr}\Bigl[a_{z}\left(\tau a_{z} +\bar{\tau}a_{\bar{z}}\right)-\bar{a}_{\bar{z}}\left(\tau\bar{a}_{z} + \bar{\tau}\bar{a}_{\bar{z}}\right)\Bigr].
\end{align}

\noindent We remark that the holomorphic formulation of the bulk theory reviewed here has been shown to be in good agreement with independent CFT calculations \cite{Kraus:2011ds,Gaberdiel:2012yb,Hijano:2013fja,Gaberdiel:2013jca}. However, as we have discussed,  the resulting black hole entropy differs from that computed with canonical methods. We will now turn our attention to a formulation of the theory that utilizes a set of canonical definitions that generalizes those of the spin-$3$ case discussed in section \ref{sec:spin3}.

\subsection{Canonical charges and conjugate potentials}
In this subsection we propose a gauge-invariant definition of the conserved charges and conjugate potentials that is consistent with the canonical structure of the theory. As we discussed in detail in the spin-2 and spin-3 examples, the starting observation is that in the presence of higher spin deformations the canonical charges are encoded in the holonomies of the connections along the $\varphi$ circle. Therefore, there exists a (constant) gauge transformation $U$ such that
\begin{align}\label{Euc DS connections}
a_{\varphi} =a_{z} + a_{\bar{z}}=
U\left(L_1+\tilde{Q}\right) U^{-1}
\end{align}
\noindent where  the matrix $\tilde Q$ is linear in the higher spin charges and satisfies
\begin{equation}
\left[L_{-1},\tilde{Q}\right] =0  \,.\label{defQQ}
\end{equation}
In other words, $\tilde{Q}$ is a highest-weight matrix that is linear in the charges. It is natural to work in highest-weight gauge for $a_\varphi\,$,
\bea\label{DS form of aphi}
\tilde a_{\varphi} = U^{-1} a_\varphi U = L_1 +\tilde Q\,,
\eea
as we did in section \ref{sec2}, see \eqref{chgtv}, but we will keep the gauge arbitrary here for easier comparison with the highest-weight gauge for $a_z\,$ that we used in the discussion of the holomorphic formalism. The definition \eqref{Euc DS connections} is to be contrasted with the holomorphic definition of charges summarized in last subsection, where the charges are introduced in the $a_{z}$ component of the connection.
Here the charge matrix directly encodes all the non-trivial information on the holonomies of the connection
around the non-contractible $\varphi$ circle.

Once we have identified the highest-weight matrix that contains the charges, the rest of the analysis is similar to that of the last subsection. In a effort of clarity for the reader, we just repeat here what we did before in terms of the new variables.  First, we can write $\tilde{Q}$ as a sum over the highest-weight generators in each $sl(2,\mathds{R})$ multiplet appearing in the decomposition of the adjoint into $sl(2,\mathds{R})$ representations as
\bea
\tilde{Q} = \frac{1}{k_{cs}} \sum_{j=2}^\infty \frac{\tilde{Q}_j}{N_{j-1}^j (\lambda)} V_{-(j-1)}^j\,, \label{matrixQ2}
\eea

\noindent where we used the normalization $N^j_{j-1}(\lambda)$ defined in \eqref{Ns}.  Using the formulae in appendix \ref{appendix:generators} we obtain
\bea
\tilde{Q}_j = k_{cs} \text{Tr}\left[V_{j-1}^j \tilde{Q}\right] = k_{cs}  (4q)^{j-2}\text{Tr}\left[(L_{1})^{j-1} \tilde{Q}\right],
\eea

\noindent where the second equality follows from \eqref{slN hw generators}. We mention that a related definition of the charges was put forward in \cite{Banados:2012ue}, where the spin-$j$ charge was identified (up to a constant) with the trace $\text{Tr}\left[\left(a_{\varphi}\right)^{j}\right]$. In the present case, starting at $j=4$ we have instead expressions like
\begin{align}
\tilde{Q}_4 ={}&
 k_{cs}(4q)^{2}\, \text{Tr}\left[L_1^3 \tilde{Q}\right]
 \nonumber\\
 ={}&
  \frac{k_{cs}}{4}(4q)^{2}\left(\text{Tr}\left[a_\varphi^4\right] - \frac{3}{5\fudge}\frac{3\lambda^{2}-7}{\lambda\left(\lambda^{2}-1\right)} \left( \text{Tr}\left[a_\varphi^2\right]\right)^2 \right),
\end{align}

\noindent so our definition of the charges will differ from that of \cite{Banados:2012ue} starting at $N=4\,$.

Generically, one can express the charge $\tilde Q_j$ in terms of $a_\varphi$ as
\bea
\tilde{Q}_j= k_{cs} \frac{(4q)^{j-2}}{j} \text{Tr}\left[a_\varphi {b}_{j-1}(a_\varphi)\right], \label{charges from traces}
\eea

\noindent where the polynomials $b_{j-1}$ are defined exactly as in section \ref{subsec:holo}, with the only difference that now they depend on $a_\varphi$ rather than $a_z\,$. In particular, for the first few values of $j$ we obtain
\begin{align}
b_{1}(a_{\varphi}) ={}&
 a_\varphi\,,
 \\
b_{2}(a_\varphi) ={}&
 a_{\varphi}^{2} - \text{Tr}\left[ a_{\varphi}^{2} \right]\left(\frac{4q}{\fudge \lambda}V^{1}_{0}\right),
 \\
 b_{3}(a_\varphi) ={}&
 a_{\varphi}^{3} - \frac{3}{5\fudge}\frac{3\lambda^{2}-7}{\lambda\left(\lambda^{2}-1\right)} \left(\text{Tr}\left[ a_{\varphi}^{2} \right]\right)^2 a_{\varphi}  -\text{Tr}\left[ a_{\varphi}^{3} \right]\left(\frac{4q}{\fudge \lambda}V^{1}_{0}\right).
\end{align}

Let us now define the sources that are conjugate to the canonical charges. We observe that the set $\{b_{j-1}(a_\phi) \}_{j \geq 2}$ form a complete basis of solutions of the equations of motion for constant connections:
\be
 \left[a_\varphi, \tau a_{z} +\bar{\tau}a_{\bar{z}}\right]=0\,.
\ee
\noindent Hence, we can expand the component of the connection along the contractible cycle of the torus in terms of these polynomials, and the coefficients in this expansion are identified with the sources $\tilde{\alpha}_j$ as
\be\label{def canonical sources}
 \left(\tau a_{z} +\bar{\tau}a_{\bar{z}}\right) = \sum_{j\geq 2} (4q)^{j-2} \tilde{\alpha}_jb_{j-1}(a_\varphi)\,,
\ee
or, equivalently, as
\be
U \left(\tau a_{z} +\bar{\tau}a_{\bar{z}}\right) U^{-1}= \sum_{j\geq 2} (4q)^{j-2} \tilde{\alpha}_jb_{j-1}(\tilde a_\varphi)\,.
\ee
Note that the sources $\tilde \alpha_j$ are defined independently of the normalization of the action ($\sim k_{cs}$) as they should.

Momentarily going back to our $N=3$ example we find for example
\begin{equation}
U^{-1}( \tau a_{z} +\bar{\tau}a_{\bar{z}} )U = \left(\begin{array}{ccc}
-\frac{4\tilde{\alpha}_{3}\tilde{Q}_{2}}{3k} & \frac{2\tilde{\alpha}_{2}\tilde{Q}_{2}}{k} + \frac{4\tilde{Q}_{3}\tilde{\alpha}_{3}}{k} & \frac{2\tilde{\alpha}_{2}\tilde{Q}_{3}}{k}  + \frac{8\tilde{\alpha}_{3}\tilde{Q}_{2}^{2}}{k^2}\\
\tilde{\alpha}_{2} & \frac{8\tilde{\alpha}_{3}\tilde{Q}_{2}}{3k} &  \frac{2\tilde{\alpha}_{2}\tilde{Q}_{2}}{k} + \frac{4\tilde{Q}_{3}\tilde{\alpha}_{3}}{k} \\
2\tilde{\alpha}_{3} & \tilde{\alpha}_{2} & -\frac{4\tilde{\alpha}_{3}\tilde{Q}_{2}}{3k}
\end{array} \right),
\end{equation}
while the $N=2$ case was derived in \eqref{spin2taut}.

\noindent It is worth emphasizing the general logic at this point: the definition of these chemical potentials (sources or intensive variables) is analogous to that used in the holomorphic formalism, in the sense that in both cases the sources are identified as the lowest weights in the component of the connection along the contractible cycle (up to a gauge transformation). The crucial difference is that the canonical charges (expectation values or extensive variables) conjugate to these sources are included in the $a_{\varphi}$ component of the connection, while in the holomorphic formalism they are included in $a_{z}$ instead.

Next, we note that \eqref{def canonical sources} implies
\be \text{Tr}\left[a_\varphi\left(\tau a_{z} +\bar{\tau}a_{\bar{z}}\right)\right] ={k_{cs}^{-1}}
 \sum_{j \geq 2} j\tilde{\alpha}_{j} \tilde{Q}_{j}\,.
\ee

\noindent Repeating the manipulations that lead to \eqref{definition sources 2h}-\eqref{definition sourcesh}, in the present case we arrive at
\begin{align}
 \text{Tr}\left[L_{1}U^{-1}\left(\tau a_{z} +\bar{\tau}a_{\bar{z}}\right)U\right] ={}&{k_{cs}^{-1}} \sum_{j\geq 2}(j-1)\tilde{\alpha}_{j} \tilde{Q}_{j}\,,
 \label{definition sources 2} \\
\text{Tr}\left[Q U^{-1}\left(\tau a_{z} +\bar{\tau}a_{\bar{z}}\right)U\right] ={}&
{k_{cs}^{-1}} \sum_{j\geq 2}\tilde{\alpha}_{j} \tilde{Q}_{j}\,.\label{definition sources}
\end{align}

\noindent When restricting to the $sl(N,\mathds{R})$ theory similar expressions apply in non-principal embeddings, where the sums would now run over the appropriate spectrum and $j$ gets replaced by the conformal weight of the corresponding operators.

We also notice that the definition of charges immediately gives a definition of vacua. In the Chern-Simons context, the vacua are simply defined as solutions with trivial holonomy around the $\varphi$ circle, just like the global AdS$_{3}$ solution of the pure gravity theory. Comparing to the usual AdS$_3$ vacuum, an equivalent definition is in terms of the following charge assignments
\bea
\tilde Q_2 = -\frac{k_{cs}}{2}\text{Tr}\left[L_0L_0\right] =-\frac{k}{4}\,,\qquad \tilde Q_j = 0 \quad \forall j \geq 3\,.
\eea

As we have reviewed, gauge connections representing smooth black geometries have trivial holonomy when transported around the contractible cycle of the torus, which in the  present context implies
\bea
\text{Tr}\left[h^n\right]=\text{Tr}\left[h^n_{\text{BTZ}}\right], \quad n=2,3,\dots
\label{holoeqho2}
\eea
\noindent where $h =2\pi\left(\tau a_{z} +\bar{\tau}a_{\bar{z}}\right)$ as before, and $h_{\text{BTZ}}$ denotes the holonomy matrix evaluated in the BTZ solution (this is, with all the higher spin charges and sources turned off). The solution to these equations determines the charges as a function of the sources (or viceversa), so we have
\bea
\tilde{Q}_j=f_j\left(\left\{\tilde{\alpha}\right\}\right), \qquad j\geq 2\,,\label{holocan}
\eea

\noindent where the function $f_j$ is the same as in \eqref{holoho}. In other words, the relation between the tilded charges and tilded chemical potentials has the same functional form as the relation between the charges and chemical potentials in the holomorphic formalism. For black holes in the $\text{hs}[\lambda]$ theory with spin 3 $\mu_3,\bar\mu_3$ deformations, the explicit expression for the function $f_j$ up to eighth order in the spin-3 source was given in \cite{Kraus:2011ds}. As we have argued, in terms of our canonical charges and chemical potentials, we just need to replace the expressions in \cite{Kraus:2011ds} by their tilded counterparts. For instance, up to the sixth order in the canonical spin $3$ source $\tilde{\alpha} = -\tilde\alpha_3\,$, the spin-2 and  spin-3 charges can be expanded as\footnote{In the language of appendix \ref{appendix:hs-lambda}, the conventions used in \cite{Kraus:2011ds} amount to $q=1/4$ and $\gamma = 24/\left(\lambda(\lambda^{2}-1)\right)\,$.}
 \bea
 \tilde{Q}_2&= & \tilde{\cL} \label{holosol}\\
 &=&-{k\over4\tilde{\tau}^2}+{5k\over3\tilde{\tau}^6}\tilde{\alpha}^2-{100k\over 3\tilde{\tau}^{10}}{\lambda^2-7\over \lambda^2-4}\tilde{\alpha}^4+{5200 k\over 27 \tilde{\tau}^{14}}{5\lambda^4-85\lambda^2+377\over (\lambda^2-4)^2}\tilde{\alpha}^6
 +\cdots\nn\\
-\tilde{Q}_3&=&\tilde{\cW}\\
&=&{2k\over 3\tilde{\tau}^5}\tilde{\alpha}-{400k\over27\tilde{\tau}^{9}}{\lambda^2-7\over \lambda^2-4}\tilde{\alpha}^3+{800 k\over 9 \tilde{\tau}^{13}}{5\lambda^4-85\lambda^2+377\over (\lambda^2-4)^2}\tilde{\alpha}^5
 +\cdots \nn
 \eea
\noindent

\subsection{Boundary term and free energy}
\label{bndterm}
Having identified the canonical charges and their conjugate sources, we will now construct a variational principle appropriate to the Dirichlet problem with fixed sources. In this section we will work directly in the highest-weight gauge for $a_{\varphi}\,$. As we shall see, physical quantities such as the free energy can be written in terms of traces which are independent of the gauge choice.

In the saddle-point approximation, valid for large temperatures and large central charges, the CFT partition function is obtained from the Euclidean on-shell action as
\begin{equation}
\ln Z = -I^{(E)}_{os} =\left. -\left(I^{(E)}_{CS} + I^{(E)}_{Bdy}\right)\right|_{os}\,,
\end{equation}

\noindent where
\begin{equation}\label{Euclidean Chern-Simons action}
I^{(E)}_{CS} =  \frac{ik_{cs}}{4\pi}\int_{M} \mbox{Tr}\Bigl[CS(A) - CS(\bar{A})\Bigr]
\end{equation}

\noindent is the Euclidean Chern-Simons action, and the boundary term $I^{(E)}_{Bdy}$ has to be constructed such that the boundary conditions (fixed sources) are enforced. Since we are going to take variations of the action, it is convenient to employ coordinates with fixed periodicity, so that the variations act on the connections (fields) only, but not on the coordinates themselves. To this end, it is useful to consider the following change of coordinates \cite{Kraus:2006wn}:
\begin{equation}
z = \frac{1-i\tau}{2}w + \frac{1+i\tau}{2}\bar{w}\,
\end{equation}

\noindent so that the boundary torus is now defined by the identifications
\begin{equation}
w \simeq w + 2\pi \simeq w + 2\pi i\,,
\end{equation}

\noindent and the components of the connection corresponding to the non-contractible and contractible cycles of the torus are given respectively by
\begin{equation}\label{zw dictionary}
a_{z} + a_{\bar{z}} = a_{w} + a_{\bar{w}}\,,\qquad \tau a_{z} + \bar{\tau}a_{\bar{z}} = i\left(a_{w}-a_{\bar{w}}\right).
\end{equation}

\noindent The volume elements are related by
\begin{equation}\label{volume elements}
i\,dw\wedge d\bar{w}=\frac{2}{(\bar{\tau}-\tau)}dz\wedge d\bar{z} = i\frac{dz\wedge d\bar{z} }{\text{Im}\left(\tau\right)} = 2\frac{d^{2}z}{\text{Im}(\tau)}\,,
\end{equation}

\noindent where $d^{2}z$ is the standard measure on the Euclidean plane.

Consider now the following boundary term:
\begin{align}\label{boundary term}
I^{(E)}_{Bdy} = \frac{ik_{cs}}{8\pi}\int_{\partial M}dw\wedge d\bar{w}\,\text{Tr}\Bigl[&
\left(a_{w}+a_{\bar{w}}-2L_{1}\right)\left(a_{w}-a_{\bar{w}}\right)\nonumber\\
&
-\left(\bar{a}_{w}+\bar{a}_{\bar{w}}-2L_{-1}\right)\left(\bar{a}_{w}-\bar{a}_{\bar{w}}\right)\Bigr].
\end{align}

\noindent Equivalently, in terms of the $(z,\bar{z})$ coordinates it reads as
\begin{align}\label{boundary term z}
I^{(E)}_{Bdy} = -\frac{ik_{cs}}{4\pi}\int_{\partial M}\frac{d^{2}z}{\text{Im}(\tau)}\text{Tr}\Bigl[&
\left(a_{z}+a_{\bar{z}}-2L_{1}\right)\left( \tau a_{z} + \bar{\tau}a_{\bar{z}} \right)
\nonumber\\
&-\left(\bar{a}_{z}+\bar{a}_{\bar{z}}-2L_{-1}\right)\left( \tau \bar{a}_{z} + \bar{\tau}\bar{a}_{\bar{z}} \right)\Bigr].
\end{align}

\noindent Writing the on-shell variation of the Euclidean Chern-Simons action \eqref{Euclidean Chern-Simons action} in the $(w,\bar{w})$ coordinates we find
\begin{align}
\left. \delta I^{(E)}_{CS}\right|_{os}
={}&
- \frac{ik_{cs}}{4\pi}\int_{\partial M}\text{Tr}\Bigl[a\wedge \delta a - \bar{a}\wedge \delta \bar{a}\Bigr]
\\
={}&
 - \frac{ik_{cs}}{4\pi}\int_{\partial M}dw\wedge d\bar{w}\,\text{Tr}\Bigl[a_{w}\delta a_{\bar{w}} - a_{\bar{w}}\delta a_{w} -\text{barred}\Bigr].
\end{align}

\noindent Combining this result with the variation of the boundary term \eqref{boundary term} we find that the variation of the full action, evaluated on-shell, is
\begin{equation}\label{variation 1}
\left. \delta I^{(E)}\right|_{os} =  \frac{ik_{cs}}{4\pi}\int_{\partial M}dw\wedge d\bar{w}\,\text{Tr}\Bigl[\left(a_{w} + a_{\bar{w}} - L_{1}\right)\delta \left(a_{w}-a_{\bar{w}}\right)-\text{barred}\Bigr].
\end{equation}

\noindent As a first check, let us explicitly show that this variation is indeed the one corresponding to fixed sources. Going back to the $(z,\bar{z})$ coordinates using \eqref{zw dictionary} we have that \eqref{variation 1} is
\begin{equation}\label{variation 2}
\left. \delta I^{(E)}\right|_{os} =  -2\pi ik_{cs}\int_{\partial M}\frac{d^{2}z}{4\pi^{2}\text{Im}(\tau)}\,\text{Tr}\Bigl[\left(a_{z} + a_{\bar{z}} - L_{1}\right)\delta \left(\tau a_{z} + \bar{\tau}a_{\bar{z}}\right)-\text{barred}\Bigr].
\end{equation}

\noindent We now plug \eqref{Euc DS connections} into this expression and obtain
\begin{equation}
\left. \delta I^{(E)}\right|_{os} =  -2\pi i k_{cs} \int_{\partial M}\frac{d^{2}z}{4\pi^{2}\text{Im}(\tau)}\,\text{Tr}\Bigl[\tilde{Q}\,\delta\left(\tau a_{z} + \bar{\tau}a_{\bar{z}}\right)-\text{barred}\Bigr].
\end{equation}

\noindent From \eqref{definition sources} it then follows that
\begin{equation}\label{checking Dirichlet variation}
\left. \delta I^{(E)}\right|_{os} =  -2\pi i \int_{\partial M}\frac{d^{2}z}{4\pi^{2}\text{Im}(\tau)}\sum_{j\geq 2} \Bigl(\tilde{Q}_{j}\delta \tilde{\alpha}_{j}-\text{barred}\Bigr),
\end{equation}

\noindent showing that the boundary term \eqref{boundary term z} is the correct one.

We emphasize that the result holds even when the solutions are non-constant: even though the polynomials $b_{j-1}(a_{\varphi})$ are not a basis of solutions for non-constant connections (so that \eqref{def canonical sources} would not hold in that case), the lowest/highest-weight structure of the solutions is still the same for spacetime-dependent connections, and as a consequence \eqref{checking Dirichlet variation} still holds given our choice of boundary term. Now, when the connections are constant we can go one step further and evaluate the on-shell action explicitly. The evaluation of the bulk Chern-Simons piece is subtle, but in \cite{Banados:2012ue} it was shown how to perform it. In the present case we obtain
\begin{align}
\left. I^{(E)}_{CS}\right|_{os}  ={}&
 -\pi k_{cs}\text{Tr}\Bigl[\left(a_{w} + a_{\bar{w}}\right)\left(a_{w}-a_{\bar{w}}\right) -\left(\bar{a}_{w} + \bar{a}_{\bar{w}}\right)\left(\bar{a}_{w}-\bar{a}_{\bar{w}}\right)\Bigr]
 \nonumber\\
 ={}&
  i\pi k_{cs}\text{Tr}\Bigl[\left(a_{z} + a_{\bar{z}}\right)\left(\tau a_{z} + \bar{\tau}a_{\bar{z}}\right)-\left(\bar{a}_{z} + \bar{a}_{\bar{z}}\right)\left(\tau \bar{a}_{z} + \bar{\tau}\bar{a}_{\bar{z}}\right)\Bigr].
\end{align}

\noindent Adding the boundary term \eqref{boundary term} evaluated for a constant connection, we find that the full on-shell action is
\begin{equation}
\left. I^{(E)}\right|_{os} = 2\pi i k_{cs} \text{Tr}\Bigl[L_{1}\left(\tau a_{z} + \bar{\tau}a_{\bar{z}}\right)  - L_{-1}\left(\tau \bar{a}_{z} + \bar{\tau}\bar{a}_{\bar{z}}\right)\Bigr].
\end{equation}

\noindent Using $\ln Z = -I^{(E)}|_{os}$ we then find the free energy as
\begin{equation}\label{free energy}
-\beta F = \ln Z  = -2\pi i k_{cs} \text{Tr}\Bigl[L_{1}\left(\tau a_{z} + \bar{\tau}a_{\bar{z}}\right) - L_{-1}\left(\tau \bar{a}_{z} + \bar{\tau}\bar{a}_{\bar{z}}\right)\Bigr].
\end{equation}

\noindent Notice that this formula depends on the connection only, and it is invariant under gauge transformations that preserve the form \eqref{DS form of aphi} of $a_{\varphi}\,$. In the $\lambda=N$ case, \eqref{free energy} gives the free energy for any $N$ and any choice of $sl(2,\mathds{R})$ embedding provided the generator $L_{1}$ is chosen appropriately.

As a second consistency check, plugging \eqref{definition sources 2} into \eqref{free energy} we find explicitly
\begin{equation}
-\beta F = \ln Z= -2\pi i \sum_{j\geq 2}\left(j-1\right)\left(\tilde{\alpha}_{j} \tilde{Q}_{j}- \text{barred}\right) \label{zbh}
\end{equation}

\noindent which is the right form obtained from dimensional analysis, in terms of the new charges and new sources.

For black holes solutions in the $\text{hs}[\lambda]$ theory deformed by a spin $3$ source $\tilde{\alpha}$ we have
\bea
\tilde{\alpha}_2\equiv\tilde{\tau}\,,\quad\tilde{\alpha}_3\equiv -\tilde{\alpha}\,
,\quad\tilde{\alpha}_j=0\,, \quad j>3\,.
\eea
 Plugging the solution to the holonomy condition \eqref{holosol} into the free energy \eqref{zbh}, we get  the free energy as a function of chemical potentials
\begin{equation}
\ln Z(\tilde{\tau},\tilde{\alpha})=
{i\pi k\over 2\tilde{\tau}}
\left[1-{4\over3}{{\tilde{\alpha}}^2\over \tilde{\tau}^4}+{400\over27}{\lambda^2-7\over \lambda^2-4}{\tilde{\alpha}^4\over \tilde{\tau}^8}-{1600\over27}{5\lambda^2-85\lambda^2+377 \over (\lambda^2-4)^2}{\tilde{\alpha}^{6}\over \tilde{\tau}^{12}}
 +\cdots\right]\label{ZZ}
\end{equation}
which reproduces the result of \cite{Kraus:2011ds} when the original variables are replaced by the tilded ones.

\subsection{Canonical entropy}

We can define the canonical entropy as the conserved charge associated with transformations generated using the component of the connection along the thermal Euclidean circle. This definition depends in a canonical fashion on the bulk action and the properties of the black hole only, just as Wald's formula \cite{Wald:1993nt}. Since this definition is expressed in terms of the connection only, it is valid for any gauge algebra and any embedding, and it is gauge-invariant.  More precisely, the gauge parameters are given by $\lambda =- ih$ and $\bar\lambda =-i\bar h\,$, where
\begin{equation}
h = 2\pi\left(\tau a_z + \bar \tau a_{\bar z}\right),\qquad \bar{h} = 2\pi\left(\tau \bar{a}_z + \bar{\tau} \bar{a}_{\bar z}\right).
\end{equation}
We showed in section \ref{sec:3} that for the spin-3 case these gauge parameters can be obtained from the asymptotic symmetry algebra analysis. The asymptotic symmetry analysis remains to be performed for the generic case. Here, we will content ourselves to define the entropy from this gauge symmetry, check that it agrees with previously obtained canonical definitions, with the Legendre transformation of the free energy, and also check that the first law is obeyed.

The infinitesimal charge for Chern-Simons theory is \cite{Regge:1974zd}
\bea
\sdelta \cQ_{\lambda,\bar\lambda}  &=& \frac{k_{cs}}{2\pi} \int_0^{2\pi} d\varphi\, \text{Tr}\left[\delta a_\varphi\, \lambda - \delta \bar a_\varphi\, \bar{\lambda}\right], \label{infch}
\eea
where the thermodynamic notation $\sdelta$ emphasizes that the quantity is not necessarily $\delta$-exact. Since  $[a_\varphi,h]=0$ from the equations of motion and since smoothness of the solution implies that there exists a group element $u$ such that $h = u^{-1}(2\pi i L_0) u\,$, one has \cite{deBoer:2013gz}
\bea
\text{Tr}\left[a_\varphi \delta h\right] = \text{Tr}\left[\bar a_\varphi \delta \bar{h}\right]  = 0\,.\label{propdeBJ}
\eea
Therefore, the charge \eqref{infch} is integrable ($\sdelta \cQ_\Lambda = \delta S$ is a total differential) and its integral is given by the entropy
\bea
S = - i k_{cs} \text{Tr}\left[ a_\varphi h -\bar a_\varphi \bar{h}\right]. \label{SdeBJg}
\eea
This general expression for the canonical entropy in terms of the connection was first found in \cite{deBoer:2013gz} by performing a Legendre transform of the free energy obtained from the on-shell value of the bulk action supplemented with boundary terms. The boundary conditions used in \cite{deBoer:2013gz} were such that the definition of the higher spin charges was the same as that employed in the holomorphic formalism, while their conjugate sources in the Euclidean setup were determined via analytic continuation of the Lorentzian variational principle appropriate for fixed deformation parameters $\mu_{j},\bar{\mu}_{j}\,$. Here, we found the free energy \eqref{free energy} adapted to the variational problem with fixed chemical potentials $\tilde\alpha_j,\bar{\tilde{\alpha}}_j\,$. By consistency, the canonical entropy should be given by the Legendre transform of the free energy under these boundary conditions. From \eqref{variation 2} we can immediately read off the term  that implements the Legendre transformation:
\begin{equation}
S = \ln Z -2\pi ik_{cs}\text{Tr}\Bigl[\left(a_{z} + a_{\bar{z}} - L_{1}\right)\left(\tau a_{z} + \bar{\tau}a_{\bar{z}}\right)-\left(\bar{a}_{z} + \bar{a}_{\bar{z}} - L_{-1}\right)\left(\tau \bar{a}_{z} + \bar{\tau}\bar{a}_{\bar{z}}\right)\Bigr]
\end{equation}

\noindent and using \eqref{free energy} we recover \eqref{SdeBJg}.

 As a further check, let us explicitly derive the first law. Using the holonomy conditions, from \eqref{propdeBJ} we immediately know that
\begin{align}
\delta S
={}&
  -2\pi i k_{cs}\text{Tr}\Bigl[\left(\tau a_{z} + \bar{\tau}a_{\bar{z}}\right)\delta \left(a_{z} + a_{\bar{z}}\right)\, - \text{barred}\Bigr]
  \nonumber\\
  ={}&
    -2\pi i k_{cs}\text{Tr}\Bigl[\left(\tau a_{z} + \bar{\tau}a_{\bar{z}}\right)\delta \left(a_{z} + a_{\bar{z}} - L_{1}\right)\, - \text{barred}\Bigr]
     \nonumber\\
  ={}&
    -2\pi i k_{cs} \text{Tr}\Bigl[\left(\tau a_{z} + \bar{\tau}a_{\bar{z}}\right)\delta \tilde{Q}\, - \text{barred}\Bigr]
\end{align}

\noindent and using \eqref{definition sources} we then conclude
\begin{equation}
\delta S
=   -2\pi i \sum_{j \geq 2}\left(\tilde{\alpha}_{j}\delta \tilde{Q}_{j}- \text{barred}\right)
\end{equation}

\noindent which explicitly shows that the first law is indeed satisfied. As a final check, we evaluate the entropy explicitly using \eqref{definition sources 2}-\eqref{definition sources} to obtain
\begin{equation}
S =  -2\pi i  \sum_{j \geq 2}j\left(\tilde{\alpha}_{j}\tilde{Q}_{j}- \text{barred}\right)
\end{equation}

\noindent which also has the right form from dimensional analysis.

In the $\lambda = N$ case, it was shown in \cite{deBoer:2013gz} that the canonical entropy can be written quite compactly as
\begin{equation}\label{dBJ entropy v2}
S = 2\pi  k_{cs}\text{Tr}\Bigl[\left( \lambda_{\varphi}- \bar{\lambda}_{\varphi}\right)L_{0}\Bigr],
\end{equation}
\noindent where the trace is taken in the $N$-dimensional representation, and $\lambda_{\varphi}$ and $\bar{\lambda}_{\varphi}$ are diagonal matrices whose entries contain the eigenvalues of $a_\varphi$ and $\bar{a}_{\varphi}\,$. We now recall that the entropy is a function of the charges (extensive variables), while the free energy is a function of the sources (intensive variables). Given that the matrices $a_{\varphi}$ and $\bar{a}_{\varphi}$ depend on the charges exclusively (and not on the sources), we see that, quite conveniently, \eqref{dBJ entropy v2} immediately provides the entropy as a function of the charges, without the need to solve the holonomy conditions that allow to express the chemical potentials in terms of the charges (or viceversa).

\subsection{Matching to the (deformed) CFT}

In \cite{Gaberdiel:2012yb}, a CFT computation for the spin three $\text{hs}[\lambda]$ black holes was carried out up to sixth order in the chemical potential,
\begin{align} \label{zcft}
\ln Z_{\text{CFT}}
 \equiv {}&
\ln\text{Tr}_{\mathcal{H}}\left[
 e^{2\pi i\left(\tau_{\text{CFT}}\hat{\cL}_{0;\text{CFT}}+\alpha_{\text{CFT}}\hat{\cW}_{0;\text{CFT}}\right)} \right] \\
={}&
{i\pi c\over 12\tau_{\text{CFT}}}\left[1-{4\over3}{{\alpha_{\text{CFT}}}^2\over \tau_{\text{CFT}}^4}+{400\over27}{\lambda^2-7\over \lambda^2-4}{\alpha_{\text{CFT}}^4\over \tau_{\text{CFT}}^8}
-{1600\over27}{5\lambda^2-85\lambda^2+377 \over (\lambda^2-4)^2}{\alpha^{6}_{\text{CFT}}\over \tau_{\text{CFT}}^{12}}+\cdots\right]\nn
\end{align}

where the conventions $q=1/4$, $\gamma = 24/(\lambda(\lambda^2-1))$ were used.

As was shown in \cite{Compere:2013aa}, the tilded charges are the zero modes of the asymptotic symmetry generators,
which suggests that the tilded variables should corresponds to the dual CFT symmetry generators.  More precisely, the dictionary from bulk to the boundary CFT is the following
\bea
6k&=&c\\
\tilde{Q}_2&=&\tilde{\cL} \rightarrow \cL_{0;\text{CFT}}\\
\tilde{Q}_3&=&-\tilde{\cW}\rightarrow -\cW_{0;\text{CFT}}\\
\tilde{Q}_j&\rightarrow& Q_{0;\text{CFT}},\quad j>3\,.
\eea
Since charges are related to chemical potential by the smoothness condition, the dictionary for the chemical potentials follows
\bea
\tilde{\alpha}_j&\rightarrow& \alpha_{j;\text{CFT}}\,.
\eea
Under the above dictionary, the free energy of the black hole \eqref{ZZ} is immediately matched to the logarithm of the CFT partition function \eqref{zcft}. We expect this matching to hold for any higher spin black hole.

\section*{Acknowledgments}

We are grateful to A.~Castro, J. de Boer, M. Gaberdiel, P. Kraus and  E. Perlmutter  for stimulating discussions and useful comments. We also heartfully thank the Centro de Ciencias Pedro Pascual in Benasque for providing a great environment during part of this work. This work is partly supported by NSF grant 1205550. W.S. is supported in part by the Harvard Society of Fellows and by the U.S. Department of Energy under grant number DE-FG02-91ER40671. G.C. is a Research Associate of the Fonds de la Recherche Scientifique F.R.S.-FNRS (Belgium).  The work of J.I.J. is supported by the research programme of the Foundation for Fundamental Research on Matter (FOM), which is part of the Netherlands Organization for Scientific Research (NWO).

\appendix

\section{Conventions}\label{appendix:generators}

\subsection{$sl(2,\mathds{R})$ algebra}\label{subsec:sl2 conventions}
The $so(2,1)$ algebra is
\begin{equation}
\left[J_{a},J_{b}\right] = \epsilon_{abc}J^{c}\,,
\end{equation}
\noindent where $J^{a} \equiv \eta^{ab}J_{b}$ and $\epsilon_{012} =-1\,$. The generators $\{L_0,L_{\pm}\}$ defined through
\begin{equation}\label{def sl2 generators}
J_{0} = \frac{L_{+} + L_{-}}{2}\,,\quad J_{1} =  \frac{L_{+} -L_{-}}{2}\,,\quad J_{2} = L_{0}\,,
\end{equation}
\noindent satisfy the $sl(2,\mathds{R}) \simeq so(2,1)$ algebra
\begin{equation*}
\left[L_{\pm},L_{0}\right] = \pm L_{\pm}\,,\,\,  \left[L_{+},L_{-}\right] = 2L_{0}\, .
\end{equation*}

\noindent We employ the usual two-dimensional representation of $sl(2,\mathds{R})$ in terms of matrices
\begin{equation}\label{2d representation of sl2r}
L_{0}=
\frac{1}{2}\left(\begin{array}{cc}
1 & 0  \\
0 & -1
\end{array}
\right),\qquad
L_{+}=
\left(\begin{array}{cc}
0 & 0  \\
1 & 0
\end{array}
\right),
\qquad
L_{-}=
\left(\begin{array}{cc}
0 & -1  \\
0& 0
\end{array}
\right).
\end{equation}
\noindent Note in particular that $\text{Tr}\left[L_0L_0\right] = 1/2\,$. The $so(2,1)$ generators in this representation are $J_{0} = -i\sigma^{y}/2$, $J_{1}=\sigma^{x}/2$, $J_{2}=\sigma^{z}/2$, where the $\sigma$'s are the Pauli matrices.

\subsection{$sl(3,\mathds{R})$ algebra}
\label{app:3}
We can parameterize the $sl(3,\mathds{R})$ algebra in terms of the generators $\{L_{0},L_{\pm}\}$ plus five $W_{j}$ generators ($j = -2, -1,0,1,2$) that transform as a spin-$2$ multiplet under the triplet $\{L_0$, $L_{\pm 1}\}$, with commutation relations
\begin{align*}
\left[L_{j},L_{k}\right]
&=
 (j-k)L_{j+k}
\nonumber\\
\left[L_{j}, W_{m}\right] &=
 (2j - m)W_{j+m}
\\
\left[W_{m},W_{n}\right] &=
 -\frac{1}{3}\left(m-n\right)\left(2m^{2} + 2n^{2} -mn-8\right)L_{m+n}\, .
\end{align*}

\noindent With this parameterization the principal and diagonal embeddings correspond to identifying the $sl(2,\mathds{R})$ generators as
\begin{align}
\text{principal embedding:} \qquad &\left\{L_{0}, L_{\pm 1}\right\}
\\
\text{diagonal embedding:} \qquad & \left\{\frac{1}{2}L_{0}, \pm\frac{1}{4}W_{\pm 2}\right\}
\end{align}

Following the conventions in \cite{Campoleoni:2010zq,Gutperle:2011kf}, in the three-dimensional defining representation of $sl(3,\mathds{R})$ (often called fundamental representation) we use the following matrix realization of the generators
\begin{align*}
L_{0} &=
\left(\begin{array}{ccc}
1 & 0 & 0 \\
0 & 0 & 0 \\
0& 0 & -1
\end{array}
\right),&
 L_{1} &=
\left(\begin{array}{ccc}
0 & 0 & 0 \\
1 & 0 & 0 \\
0& 1 & 0
\end{array}
\right),&
L_{-1} &=
-2\left(\begin{array}{ccc}
0 & 1 & 0 \\
0& 0 & 1\\
0 & 0 & 0
\end{array}
\right),
\label{3d representation of sl2r}
\\
W_{2} &=
 2
\left(
\begin{array}{ccc}
0& 0 & 0 \\
0 & 0 & 0\\
1& 0 & 0
\end{array}
\right),&
W_{1} &=
\left(
\begin{array}{ccc}
0& 0 & 0 \\
1 & 0 & 0\\
0& -1 & 0
\end{array}
\right),&
W_{-2} &=
8
\left(
\begin{array}{ccc}
0& 0 & 1 \\
0 & 0 & 0\\
0& 0 & 0
\end{array}
\right),&
\nonumber\\
W_{-1} &=
2
\left(
\begin{array}{ccc}
0& -1 & 0 \\
0 & 0 & 1\\
0& 0 & 0
\end{array}
\right),&
W_{0} &= \frac{2}{3}
\left(
\begin{array}{ccc}
1& 0 & 0 \\
0 & -2 & 0\\
0& 0 & 1
\end{array}
\right).
\end{align*}
\noindent In the above representation, the non-zero traces are given by
\begin{gather*}
\begin{align}
\mbox{Tr}\left[L_{0}L_{0}\right] &=2\,,&
 \mbox{Tr}\left[L_{1}L_{-1}\right] &=-4\,,
\nonumber\\
 \mbox{Tr}\left[W_{1}W_{-1}\right] &= -4\,,&
  \mbox{Tr}\left[W_{2}W_{-2}\right] &= 16\,,
\end{align}
 \\
  \mbox{Tr}\left[W_{0}W_{0}\right] = \frac{8}{3}\, .
\end{gather*}
We denote the identity matrix as $\mathds{1}$ with trace $\text{Tr}\left[\mathds{1}\right]=3$.

\subsection{$\text{hs}[\lambda]$ algebra}\label{appendix:hs-lambda}

The higher spin algebra $\text{hs}[\lambda]$ has generators
\bea
V_n^s\,,\qquad s \geq 2\,,\qquad |n|<s\,,
\eea
We define as usual the lowest order generators as
\bea
L_m = V^2_m\,,\qquad W_m = V^3_m\,,\qquad J_m = V^4_m\,.
\eea
The commutators are
\bea
 [L_m,V_n^s] &=& (-n+m(s-1))V^s_{m+n}\,,\\
\mbox{} [V_m^s,V_n^t\mbox{}]&=&\sum_{u=2, \text{even}}^{s+t-1}g_u^{st}(m,n ; \lambda)V^{s+t-u}_{m+n}
\eea
with the following structure constants
\bea
g_u^{st}(m,n ; \lambda) &=& \frac{q^{u-2}}{2(u-1)!}\phi_u^{st}(\lambda) N^{st}_u(m,n)\,,\\
N^{st}_u(m,n) &=& \sum_{k=0}^{u-1}(-1)^k \left( \begin{array}{c} u-1\\ k \end{array} \right) [s-1+m]_{u-1-k} [s-1-m]_k\nn\\
&&  [t-1+n]_k [t-1-n]_{u-1-k}\\
\phi_u^{st}(\lambda) &=& \mbox{}_4 F_3 \left[ \begin{array}{c} \frac{1}{2}+\lambda ,\frac{1}{2}-\lambda,\frac{2-u}{2},\frac{1-u}{2} \\ \frac{3}{2}-s,\frac{3}{2}-t,\frac{1}{2}+s+t-u \end{array} \Big| 1 \right]
\eea
where $[a]_n \equiv \Gamma(a+1)/\Gamma (a+1-n)$ is the descending Pochhammer symbol. These definition match with the ones of \cite{Kraus:2011ds}. The conventions of \cite{Gaberdiel:2011wb} are obtained after the substitution $V_n^s \rightarrow V_n^s/4 $. The number $q$ is arbitrary and can be scaled away by taking $V_n^s \rightarrow q^{s-2} V_n^s $.

If $\lambda = N$ with any integer $N\geq 2$, an ideal $\chi_N$ appears, consisting of all generators $V^s_n$ with $s > N$. Factoring over this ideal truncates to the finite algebra $\text{sl}(N)$,
\bea
\text{sl}(N) = \frac{\text{hs}(N)}{\chi_N} \qquad (N \geq 2).
\eea
The relation
\bea
\phi_u^{st}\left( \frac{1}{2}\right) = \phi_2^{st}(\lambda) = 1\,,
\eea
implies the isomorphism $\text{hs}\left[\frac{1}{2}\right] \sim \text{hs}(1,1)$. The property $N_u^{st}(0,0) = 0$ implies that all zero modes $V^s_0$ commute.

The lone star product is defined on $\text{hs}[\lambda] \oplus \mathds{C}$ as
\bea
\mbox{} V_m^s \star V_n^t\mbox{}&=&\frac{1}{2}\sum_{u=1,2,3}^{s+t-1}g_u^{st}(m,n ; \lambda)V^{s+t-u}_{m+n}\,.
\eea
One can in fact restrict the upper bound in the sum to $s+t-1-|s-t|$ since the rest vanishes. The additional generator $V^1_0$ is central and formally equal to the identity. We will omit the explicit $\star$ symbol in the main text when considering products of generators. Powers of the generators are defined using the star product.

We define the trace of the generators as
\bea
\text{Tr}\left[V_0^1\right] = \fudge \frac{\lambda}{4q}\,, \qquad \text{Tr}\left[V_m^s\right] = 0\,, \quad \forall (s,m)\neq (1,0)\,,
\label{deftr}
\eea
where $\gamma$ is an arbitrary parameter which encodes the different possible conventions for the trace. An invariant symmetric bilinear trace can then be defined as
\bea
\text{Tr}[V_m^s,V_n^t]= \text{Tr}\left[V^s_m \star V^t_n\right] .
\eea
The trace is explicitly
\begin{align}
\text{Tr}\left[V_m^s , V_n^t \right] ={}&
 N^s_m(\lambda) \delta^{st}\delta_{m,-n}\,,\label{Ns}\\
\text{with}\qquad N^s_m(\lambda) \equiv{}&
 \fudge \frac{\lambda}{8q} g^{ss}_{2s-1}(m,-m;\lambda)\,. \label{hs lambda norm}
\end{align}

The $sl(3,\mathds{R})$ generators written down in appendix \ref{app:3} are consistent with the choice $q=1/2$ and $\fudge=1\,$. The identity matrix is $\mathds{1} = 2 V_0^1$ and the star product reduces to matrix multiplication. When $\fudge=1$ and $q=1/4$, the normalization agrees with that of \cite{Castro:2011iw}. When $\fudge = \frac{24}{\lambda(\lambda^2-1)}$ we have $\text{Tr}\left[L_0,L_0\right]=2$ which reproduces the convention used in \cite{Gaberdiel:2012yb}. Unless explicitly noted, we will keep $\fudge$ and $q$ arbitrary so that our expressions can be easily adapted to different conventions.

The relation
\bea
N_u^{st}(m,n) = (-1)^{u+1}N_u^{ts}(n,m)
\eea
shows the consistency of the lone star product with the $\text{hs}[\lambda]$ algebra
\bea
\left[V^s_m, V^t_n\right] = V^s_m \star V^t_n -   V^t_n\star V^s_m \,.
\eea
Note that $j-1$ times the star product of $L_1$ gives
\bea
V_{j-1}^j = (4q)^{j-2} (L_1)^{j-1}\,.\label{slN hw generators}
\eea
This property matches the one discussed in  \cite{Gaberdiel:2011wb} where $q=1/4$ is used.

\providecommand{\href}[2]{#2}\begingroup\raggedright\endgroup

 \end{document}